\documentclass[aps,pre,citeautoscript,superscriptaddress,nofootinbib,preprint]{revtex4-1}  
\usepackage{amsmath,amssymb,bm,bbm} 
\usepackage{graphicx}  
\usepackage{wasysym}
\usepackage{color} 
\usepackage{subcaption}
\usepackage{relsize}
\usepackage{multirow}
\usepackage{booktabs}

\usepackage{pifont}
\newcommand{\cmark}{\ding{51}}%
\newcommand{\xmark}{\ding{55}}%
\usepackage[colorlinks=true]{hyperref}  
\hypersetup{
    bookmarks=true,         
    unicode=false,          
    pdftoolbar=true,        
    pdfmenubar=true,        
    pdffitwindow=false,     
    pdfstartview={FitH},    
    pdftitle={Triangular antiferromagnetism on the honeycomb lattice of twisted bilayer graphene},    
    pdfauthor={Alex Thomson, Shubhayu Chatterjee, Subir Sachdev and Mathias Scheurer},     
    pdfsubject={},   
    pdfcreator={},   
    pdfproducer={}, 
    pdfkeywords={} {} {}, 
    pdfnewwindow=true,      
    colorlinks=true,       
    linkcolor=magenta, 
    citecolor=blue,        
    filecolor=magenta,      
    urlcolor=blue           
} 

\usepackage{amsfonts}
\usepackage{upgreek}
\usepackage{slashed}
\usepackage{latexsym}
\usepackage{dsfont}
\usepackage{todonotes}

\newcommand{\beq}{\begin{equation}}
\newcommand{\eeq}{\end{equation}}
\def\bea{\begin{eqnarray}}
\def\eea{\end{eqnarray}}
\def \k{{\bm k}}
\def \Q{{\bm Q}}
\def \S{{\bm S}}
\def \r{{\bm r}}

\newcommand{\nn}{\nonumber \\}
\renewcommand{\vec}[1]{\boldsymbol{#1}}

\newcommand{\equref}[1]{Eq.~(\ref{#1})}

\newcommand{\secref}[1]{Sec.~\ref{#1}}
\newcommand{\figref}[1]{Fig.~\ref{#1}}
\newcommand{\refcite}[1]{Ref.~\onlinecite{#1}}
\newcommand{\refscite}[1]{Refs.~\onlinecite{#1}}

\newcommand{\appref}[1]{Appendix~\ref{#1}}

\newcommand{\pdagger}{{\phantom{\dagger}}}
\usepackage{braket}

\newcommand{\br}{{\boldsymbol{r}}}

\newcommand{\veo}{{\boldsymbol{e}_1}}
\newcommand{\vet}{{\boldsymbol{e}_2}}
\newcommand{\vK}{{\boldsymbol{K}}}
\newcommand{\vk}{{\boldsymbol{k}}}
\newcommand{\Hc}{\text{H.c.}}

\newcommand{\smHex}{{\text{\hexagon}}}
\newcommand{\smTri}{\mathsmaller{\triangle}}

\begin{document}

\title{Triangular antiferromagnetism on the honeycomb lattice of twisted bilayer graphene}

\author{Alex Thomson}
\affiliation{Department of Physics, Harvard University, Cambridge MA 02138, USA}

\author{Shubhayu Chatterjee}
\affiliation{Department of Physics, Harvard University, Cambridge MA 02138, USA}

\author{Subir Sachdev}
\affiliation{Department of Physics, Harvard University, Cambridge MA 02138, USA}
\affiliation{Perimeter Institute for Theoretical Physics, Waterloo, Ontario, Canada N2L 2Y5}

\author{Mathias S.~Scheurer}
\affiliation{Department of Physics, Harvard University, Cambridge MA 02138, USA}

\date{\today
\\
\vspace{0.4in}}

\begin{abstract}
We present the electronic band structures of states with the same symmetry as the three-sublattice planar antiferromagnetic order of the triangular lattice. 
Such states can also be defined on the honeycomb lattice provided the spin density waves lie on the bonds. We identify cases which are consistent with observations on twisted bilayer graphene: a correlated insulator with an energy gap, yielding a single doubly-degenerate Fermi surface upon hole doping. We also discuss extensions to metallic states which preserve spin rotation invariance, with fluctuating spin density waves and bulk $\mathbb{Z}_2$ topological order.
\end{abstract}
\maketitle

\section{Introduction}
Twisted graphene bilayers \cite{MacDonald11,Neto1,Mele,Neto2,MoonKoshino,Koshino} have recently been observed \cite{Pablo1,Pablo2} to exhibit
correlated insulating behavior and superconductivity for twist angles close to the magic angle. This has stimulated much theoretical interest \cite{Xu18,Po18,Fu18,Vafek18,Fu218,Senthil2,MacDonald18,roy2018unconventional,Scalettar18,Baskaran18,Phillips18,irkhin2018dirac,Dodaro18,Ma18,zhang2018low,ray2018wannier,liu2018chiral,Lee18,Finn18,fidrysiak2018unconventional,rademaker2018charge,kennes2018strong,isobe2018superconductivity,Fu218,you2018superconductivity,wu2018emergent,pizarro2018nature,zhang2018moir,wu2018theory,ochi2018possible,kennes2018strong}
on correlated electron phases on triangular and honeycomb lattices. 

While there is a significant debate on the precise nature of the needed lattice model to describe these phenomena,
experimental observations \cite{andrei11} 
clearly indicate that the electron charge density is concentrated on a moir\'e triangular
lattice. This suggests that the consequences of local correlations should be similar to those on the triangular lattice. On the other hand, symmetry and topological aspects of the band structure require that the model be formulated using the Wannier orbitals of a honeycomb lattice \cite{Po18,Fu18,Vafek18}.

In this paper, we explore the electronic band structure of states with the space group and spin-rotation symmetries of 120$^\circ$ coplanar antiferromagnetism on the triangular lattice. By transforming to a rotating reference frame \cite{SS09,CSS17,Scheurer18}, 
our results can also be applied to more exotic topologically ordered states which preserve spin rotation invariance,
as we will describe in Appendix~\ref{app:topo}. One of our main observations is that such triangular antiferromagnetic
states can also appear on the honeycomb lattice, after we allow for non-zero inter-site spin moments. The symmetry of the triangular antiferromagnet is such that all on-site spin moments on the honeycomb lattice vanish,
and so such a state will not be observable in studies \cite{Lee18,Ma18} which only examine on-site moments.

Given the variety of experimental situations, we explore the band structure in a number of different models,
all of which have two electrons per unit cell in the correlated insulator.
We start in Section~\ref{TriangularLatticeModel} by studying a triangular lattice model with two orbitals per site \cite{Xu18,Dodaro18}. 
The spins of the electrons in the two orbitals are parallel on the same site, while for each orbital the spin is antiferromagnetically ordered on the triangular lattice. 

Multiple papers assert that any tight-binding model for the bilayer graphene system should be constructed on a honeycomb lattice; in Section~\ref{HoneycombLatticeMagn} we study the 120$^\circ$ antiferromagnet from this perspective \cite{Fu18,Vafek18,Po18}.
While these works agree that the tight-binding model should have two orbitals per lattice site, Refs.~\citenum{Fu18} and~\citenum{Vafek18}  
implement a `valley rotation symmetry' in a different manner than Ref.~\citenum{Po18}.
In Section~\ref{sec:HoneycombQuarterFilling}, we consider the model of Ref.~\citenum{Fu18} and~\citenum{Vafek18} at quarter-filling.
We next study antiferromagnetism in the model of Ref.~\citenum{Po18} in the `intervalley coherent' phase in Section~\ref{sec:HoneycombHalfFilling}.
Half of the original degrees of freedom are gapped out in this phase so that the appropriate description is a honeycomb lattice with \emph{no} orbital degeneracy at half-filling \cite{Po18,Lee18}. 

Our aim is to search for cases consistent with observations in \refscite{Pablo1,Pablo2}: an energy gap in a correlated insulator, and
a single doubly-degenerate hole Fermi surface on the hole-doped side. We will show that the half-filled triangular lattice model of Section~\ref{TriangularLatticeModel}, and the quarter-filled honeycomb lattice model of Section~\ref{sec:HoneycombQuarterFilling} can display the required features. The half-filled honeycomb lattice model of Section~\ref{sec:HoneycombHalfFilling} requires the additional valence bond ordering \cite{Lee18}.
These results are summarized in Tab.~\ref{tab:ResultSum}.

\begin{table}
    \centering
    \begin{tabular}{c lc cc lc cc cc}
    \hline\hline
         &\multicolumn{1}{c}{lattice} & &  \multicolumn{1}{c}{orbitals/site}   & &  \multicolumn{1}{c}{order parameter} & & insulating& & FS structure  &\\ \hline
         &  triangular & & 2 & & on-site AF & &  \cmark  &  & \cmark &\\ \hline
         &\multirow{3}{*}{honeycomb}  &   &   \multirow{3}{*}{2}   &   &   AF on bonds, diagonal    &   &   \multirow{1.5}{*}{\xmark}  &   &   \multirow{1.5}{*}{N/A}    &\\[-8pt]
        &&   &      &   &   in orbital indices    &   &    &   &     &\\ %
        &&   &       &   &   AF on bonds, non-diagonal  &   &   \multirow{1.5}{*}{\cmark}  &   &   \multirow{1.5}{*}{\cmark}& \\[-8pt]
        &&   &       &   &   in orbital indices &   &   &   &   &\\\hline
        &\multirow{3}{*}{honeycomb} &   &   \multirow{3}{*}{1}     &   &    AF on bonds   &   &   \xmark  &   &   N/A &\\
        &&   &        &   &   KVBS &   &   \cmark  &   &   \xmark  & \\
        &&  &        &   &   AF on bonds \& KVBS &   &   \multirow{1}{*}{\cmark} &   &  \multirow{1}{*}{\cmark}  &
        \\ \hline\hline
    \end{tabular}
    \caption{Summary of results presented in this paper. In the third column, ``AF order" refers to the 120$^\circ$ coplanar antiferromagnetic order described in Sec.~\ref{TriangularLatticeModel}, while ``KVBS" refers to Kekul\'{e} valence bond solid order described in Sec.~\ref{sec:HoneycombHalfFilling}. The fourth column indicates whether an insulating state with two electrons per unit cell is possible for each model. For those models that allow for an insulator, the final column indicates whether the experimentally observed twofold degenerate Fermi surfaces can be obtained on the hole-doped side of the insulating state.}
    \label{tab:ResultSum}
\end{table}

\begin{figure}
   \centering
    \includegraphics[width=0.90\textwidth]{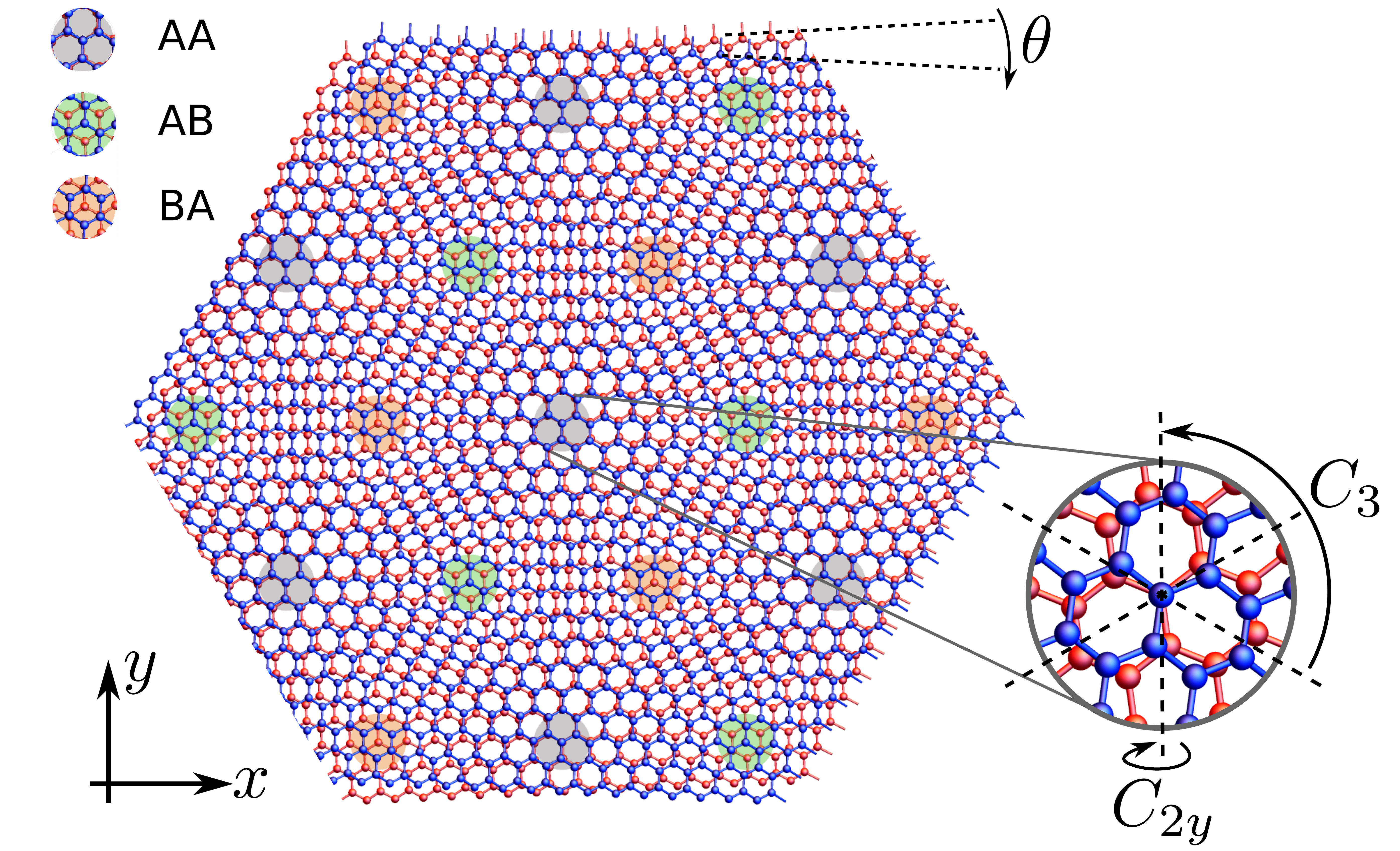}
    \caption{Moir\'e superlattice resulting from two graphene sheets, indicated in red and blue, twisted by angle $\theta$ relative to each other. Here we focus on commensurate twist angles and assume that the rotation axis goes through the AA site \cite{Fu18,Vafek18}. The regions of local AA, AB, and BA stacking each form a triangular lattice. The generators of the point group $D_3$ of the system are illustrated in the inset, where the twist angle $\theta$ has been chosen larger to make the geometry more clearly visible.}
    \label{fig:TBGLatticeGeometry}
\end{figure}

\begin{figure}
\begin{center}
\includegraphics[width=0.98\textwidth]{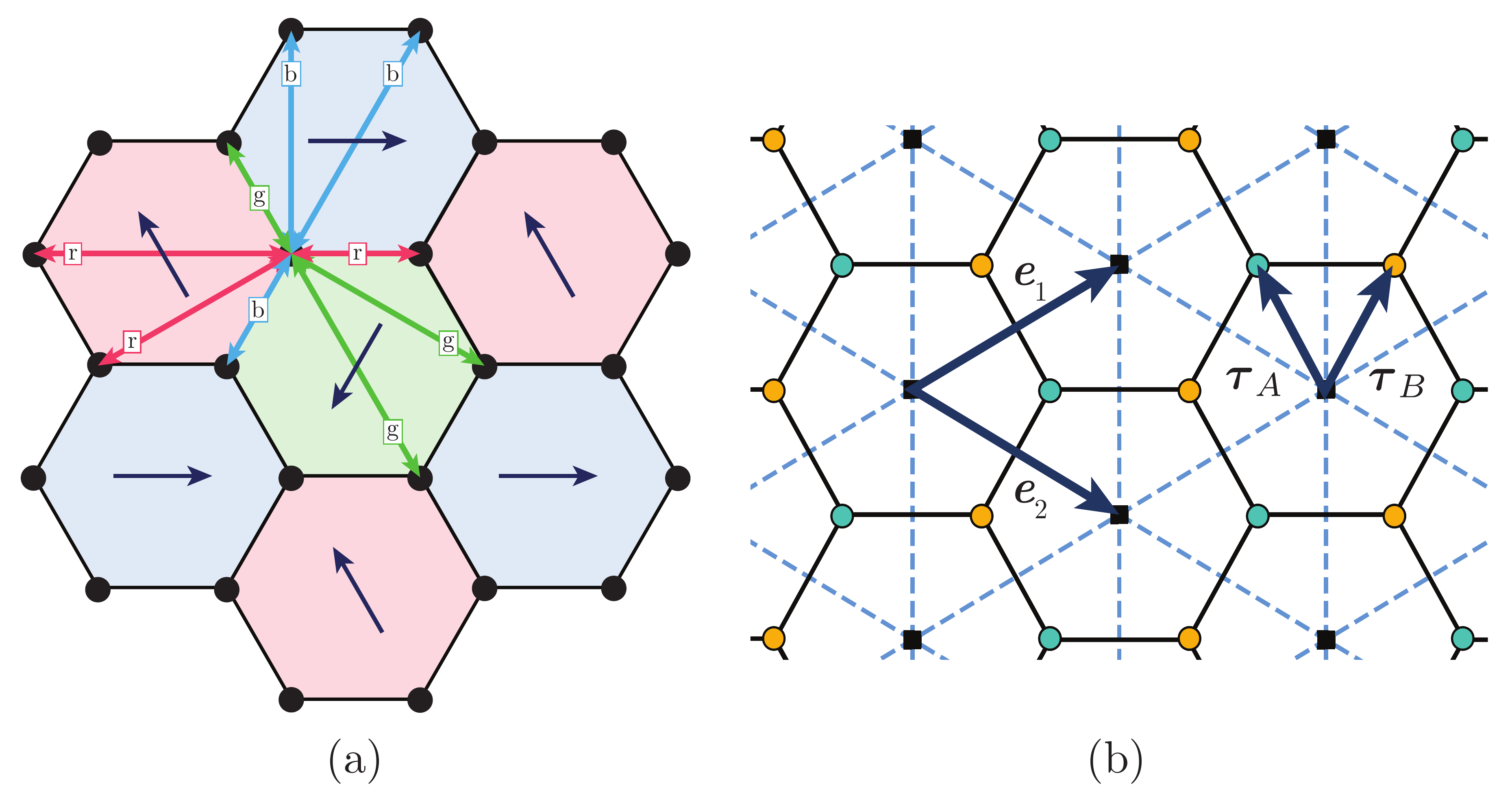}
\end{center}
\caption{
(a) Schematic of the triangular antiferromagnetism on the honeycomb lattice. The red, green, and blue (r, g, and b) hexagons  label the three spin orientations indicated by the arrows at the hexagon centers: $\vec{S}_\mathrm{red} = (-1/2,\sqrt{3}/2,0,0)$, $\vec{S}_\mathrm{green} = (-1/2,-\sqrt{3}/2,0)$,
$\vec{S}_\mathrm{blue} = (1,0,0)$. 
The spins on the red, green, or blue bonds have the same orientation as the spin of that colour.
All types of bonds considered are shown emanating from a single site.
(b) 
Triangular lattice sites are shown as squares. The dual honeycomb lattice sites are indicated by turquoise and orange circles, corresponding to the $A$ and $B$ sublattices, respectively.
The primitive vectors, $\veo$ and $\vet$, are drawn in navy, and $\vec{\tau}_{A,B}$ indicate which honeycomb lattice site is associated to each site of the triangular lattice: $\br_{A,B}=\br+\vec{\tau}_{A,B}.$
}
\label{fig:honeycomb}
\end{figure}

\section{Magnetic order on the triangular lattice}
\label{TriangularLatticeModel}
We begin by first considering a minimal, phenomenological model on the triangular lattice \cite{Xu18,Dodaro18}. 

Close to charge-neutrality, the low-energy degrees of freedom of two non-interacting graphene layers are given by Dirac fermions at the two valleys, $\vK_\mathrm{lbz}$ and $\vK'_\mathrm{lbz}$. 
Here, the subscript ``lbz" is used to emphasize that these are momenta belonging to the \emph{large} Brillouin zone (BZ) of the individual graphene layers.
(Technically, when the graphene sheets are twisted relative to one another, these momentum must be as well; since the twist angle is small, there is no difficulty in identifying which Dirac point momenta on either layer should be associated with $\vK_\mathrm{lbz}$ and $\vK'_\mathrm{lbz}$.)
The momentum transfer between the two valleys is very large and mixing is typically assumed to only occur between states originating from the same valley  \cite{MacDonald11}.  
Motivated by the experimental observation that the electronic density is concentrated in the vicinity of the AA stacking regions of the moir\'e superlattice \cite{andrei11}, which form a triangular lattice (see gray dots in \figref{fig:TBGLatticeGeometry}), Xu and Balents \cite{Xu18} introduced the Hubbard-like Hamiltonian $H^\smTri=H_t^\smTri + H_{\text{int}}^\smTri$
with degenerate valley-orbitals on each site:
\begin{subequations}
\begin{align}
H^{\smTri}_t &=- t \sum_{\braket{\vec{r},\vec{r}'}} \sum_{v} \left(c^\dagger_{v,\vec{r}} c^\pdagger_{v,\vec{r}' }  + \Hc \right), \label{HamiltonianTr0} \\ 
H_{\text{int}}^\smTri &= U \sum_{\vec{r}} \left(\sum_{v} c^\dagger_{v,\vec{r}} c^\pdagger_{v,\vec{r}} \right)^2 - V \sum_{\vec{r}} \left(\sum_{v} c^\dagger_{v,\vec{r}}\vec{\sigma}c^\pdagger_{v,\vec{r}}\right)^2. \label{HamiltonianTrInt}
\end{align}\label{HamiltonianTr}\end{subequations}
Here, $c_{v,\vec{r}}$ annihilates an electron at triangular lattice site $\br=r_1\veo+r_1\vet$, $r_{1,2}\in\mathds{Z}$ (see \figref{fig:honeycomb}(a))  and valley $v\in\{\vK_\mathrm{lbz},\vK_\mathrm{lbz}'\}$. 
An additional spin index has been suppressed.
In the noninteracting term, \equref{HamiltonianTr0}, $\braket{\vec{r},\vec{r}'}$ refers to nearest neighbors and $t>0$ will be assumed as we are interested in describing the nearly flat bands below charge neutrality. 

This model is invariant under the following spatial symmetry operations: superlattice translation $T_j$ along $\vec{e}_j$, $j=1,2$ (see Fig.~\ref{fig:honeycomb}(b)), two-fold rotation
$C_{2y}$ about the ${y}$-direction, which can alternatively be viewed as a reflection about $y$ for the two-dimensional tight-binding model, and six-fold rotation $C_{6}$ perpendicular to the graphene sheets.
These transform the Bravais lattice vector $\br$ as
\begin{align}
    T_1&: \quad(r_1,r_2)\to (r_1+1,r_2),
    &
    T_2&: \quad (r_1,r_2)\to (r_1,r_2+1),
    \notag\\
    C_{2y}&:\quad(r_1,r_2)\to (-r_2,-r_1),
    &
    C_{6}&:\quad (r_1,r_2)\to (r_1+r_2,-r_1).
\label{eq:LS}    
\end{align}
We remark that $C_{6}$ should be viewed as an approximate symmetry of the model since the microscopic twisted bilayer structure in \figref{fig:TBGLatticeGeometry} only has three-fold rotation symmetry $C_3$ \cite{Fu18,Vafek18}. 
We will come back to the issue of enhanced rotational symmetry when discussing the honeycomb-lattice models in \secref{HoneycombLatticeMagn} below.


As we alluded to in the introduction, there are several issues with using a triangular lattice model to describe the bilayer graphene system.
Notably, a model on the triangular lattice cannot give rise to the irreducible representations observed in band structure computations at the high symmetry points $\Gamma$ and $K$ in the Brillouin zone \cite{Po18,Fu18,Vafek18}.
Nevertheless, the triangular lattice model can be seen as a simple, phenomenological caricature of the system to gain physical intuition. We also note that triangular lattice models have been proposed to describe the moir\'e bands arising in twisted transition metal dichalcogenide heterobilayers \cite{wu2018hubbard}. In this work, we will use the model to motivate the symmetry of the order parameter for the correlated insulating states of twisted bilayer graphene at low twist angles. We will construct an order parameter with the same symmetries for the honeycomb-lattice models below, which is capable of reproducing the correct irreducible representations, and compare the resulting spectra in the magnetically ordered state in the different models.

\begin{figure}
\begin{center}
\includegraphics[width=0.8\textwidth]{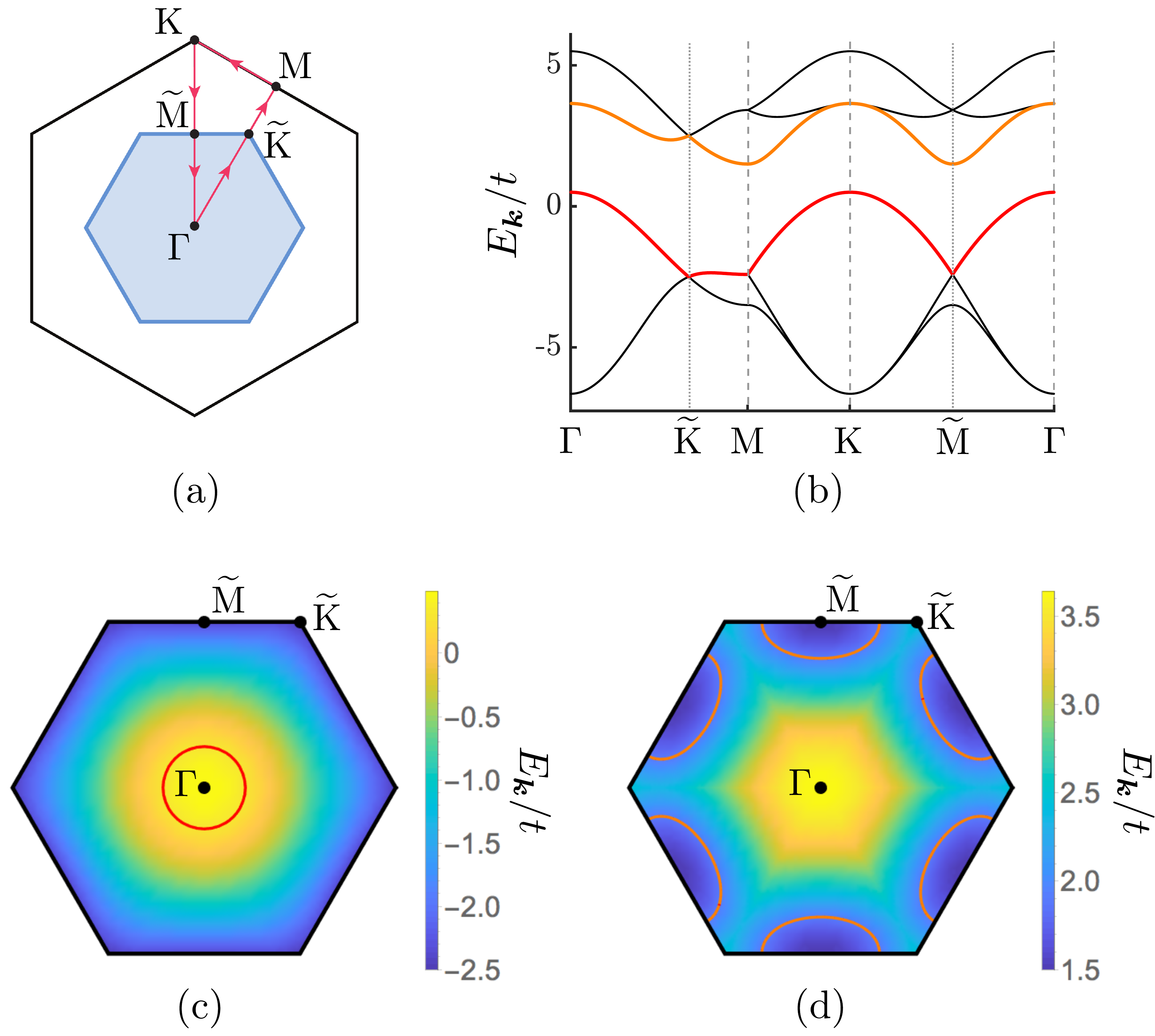}
\end{center}
\caption{The spectrum of the triangular lattice model with 120$^\circ$ coplanar spin-order, $H^{\smTri}_t+H_{\text{mag}}^\smTri$, along the one-dimensional momentum cut indicated in (a) is shown in (b). The Brillouin zone (BZ) of the moir\'e lattice is the large hexagon depicted in black. It contains the reduced BZ, colored in blue, which is $1/3$ the size of the full BZ. (c) and (d) show the Fermi surfaces (red and orange solid lines) in the magnetic BZ along with the momentum dependence of the band that crosses the Fermi energy [indicated in red and orange in part (b)] upon hole and electron doping of the half-filled state, respectively. In all plots, we have chosen $P_0/t=2.5$ leading to a full gap at half-filling of the triangular lattice, which corresponds to quarter filling of the flat-bands. 
}
\label{fig:TriangularOPAndSpectrum}
\end{figure}

The interaction term in \equref{HamiltonianTrInt}, proposed in \refcite{Xu18}, consists of a local repulsion, $U>0$, and a Hund's coupling, $V>0$. For these interactions, it is natural to expect 120$^\circ$ coplanar spin order with the spin in both valleys oriented parallel in the half-filled triangular model ({\it i.e.} at quarter filling of the nearly flat bands). 
For the insulating state, one can eliminate the charge degrees of freedom from Eq.~(\ref{HamiltonianTr}) by a Schrieffer-Wolff transformation to obtain a  Heisenberg model for the spins (in the $|t|,\,V \ll U$ limit). Analytical and numerical studies \cite{JG1989,BLP1992,Deutscher1993} have shown that such an ordered state is preferred for the nearest-neighbor Heisenberg model on the triangular lattice. Turning on a weak positive $V$ would then favor parallel orientation of spins on the two orbitals at the same site. 
At the mean-field level, this order can be represented by the term \cite{Tremblay95}
\begin{subequations}
\begin{align}\label{eqn:MagOrder}
    H_{\text{mag}}^\smTri&=P_0\sum_{\vec{r}} \bm{\mathcal{P}}(\vK\cdot\br)\cdot \sum_{v} c_{v,\br}^\dag \bm{\sigma} c^\pdagger_{v,\br},
\end{align}
where
\begin{align}\label{eqn:OrderParamVector}
    \bm{\mathcal{P}}(\theta)&=\left(\cos \theta,\sin \theta,0\right),
    &
    \vK=\frac{4\pi}{3}\veo=
    \frac{4\pi}{3}\left(
    \frac{\sqrt{3}}{2}\,\bm{\hat{x}}+\frac{1}{2}\,\bm{\hat{y}}
    \right),
\end{align}\label{AFMStateWeWant}\end{subequations}
and is illustrated in \figref{fig:honeycomb}(a).

As pointed out in \refcite{Dodaro18}, the detailed form of the interactions is unclear, even in the model on the triangular lattice. 
In particular, the extended Wannier orbitals might give rise to non-negligible further neighbor spin-exchange interactions, and phonon-mediated interactions can reduce on-site repulsion. These effects lead to geometric frustration which can, in principle, destabilize the 120$^\circ$ magnetically ordered phase and lead to a distinct magnetic ordering pattern or a quantum-disordered spin liquid phase.
Nonetheless, our assumption of 120$^\circ$ antiferromagnetic order, defined in \equref{AFMStateWeWant}, is supported by the observation that  it is the only configuration (besides ferromagnetic spin order) that preserves all the lattice symmetries in \equref{eq:LS} in all spin-rotation invariant observables, such the charge density (see \appref{app:SymCons} for details). 
In \refcite{Pablo1}, it was shown that the correlated insulating state can be suppressed by the application of a magnetic field, and the critical Zeeman coupling is estimated to be comparable to the thermal excitation gap. Since this  is incompatible with ferromagnetic spin order, we believe that the order parameter in \equref{AFMStateWeWant} is the most natural starting point.


In \figref{fig:TriangularOPAndSpectrum}, we show the spectrum and resulting Fermi surfaces upon doping of the insulating state obtained from the mean-field Hamiltonian $H_t^\smTri+H_{\text{mag}}^\smTri$. 
We first note that the order parameter in \equref{AFMStateWeWant} can induce a full gap for sufficiently large $P_0/t$ (see \figref{fig:TriangularOPAndSpectrum}(b)). 
As we will see below, this is different in the honeycomb lattice model. Secondly, in this regime, as displayed in \figref{fig:TriangularOPAndSpectrum}(c), we obtain a single, doubly degenerate, hole Fermi surface on the hole-doped side,  which is consistent with the quantum oscillations reported in \refscite{Pablo1,Pablo2}. \figref{fig:TriangularOPAndSpectrum}(d) shows, for completeness, that this model predicts three inequivalent electron pockets for the electron-doped case.  

The Fermi surface structure above generalizes to the case where the antiferromagnetic order is `quantum-fluctuating'
so that spin rotation invariance is preserved  \cite{SS09,CSS17,Scheurer18}. Such states have $\mathbb{Z}_2$ topological order (similar to that of the `toric code') co-existing with gapless fermionic excitations on Fermi surfaces like those discussed above. 
This generalization is described in more detail in Appendix~\ref{app:topo}.

\section{Magnetic order on the honeycomb lattice}\label{HoneycombLatticeMagn}

In the previous section, we used a triangular lattice model to motivate the 120$^\circ$ coplanar antiferromagnetic state as a natural candidate for the correlated insulator with two electrons per unit cell.
However, although the charge is concentrated on the points of a triangular lattice, symmetry arguments imply that a proper tight-binding model should be formulated on a honeycomb lattice \cite{Po18,Fu18,Vafek18,Fu218}.
In this section, we study the 120$^\circ$ antiferromagnet state in the context of two different honeycomb lattice models which we introduce below.

For both of these models, there exists a caveat to any study of the 120$^\circ$ antiferromagnet: this state cannot be represented as an on-site magnetic order parameter on the honeycomb lattice. 
This is readily understood through inspection of \figref{fig:honeycomb}(a); any site of the honeycomb lattice is surrounded by all three spin configurations of the 120$^\circ$ coplanar state (the three different colors), and, hence, cannot be consistently assigned one of those values without breaking the lattice symmetries.
We discuss this more rigorously in Appendix~\ref{NoLocalOrderParameter}.

However, an order parameter can be formulated so long as it lies on the bonds:
\begin{align}\label{eqn:GeneralOrderParam}
    H^{\smHex}_\mathrm{mag}&=\sum_\br \bm{\mathcal{P}}(\vK\cdot\br)\cdot \left(
    \sum_{\substack{\br',\br''\\\alpha,\beta}}
    f^{\alpha\beta}_{\bm{a},\bm{a}'}c_{\alpha,\br+\bm{a}}^\dag \bm{\sigma}c_{\beta,\br+\bm{a}'}
    \right),
\end{align}
where $\alpha,\beta$ label the sublattice and, potentially, an orbital index, and $\bm{a}$, $\bm{a}'$, $\vec{r}$ are Bravais lattice vectors.
In this paper, we limit our study to magnetic bond order located on nearest-neighbour, next-nearest neighbour and third-nearest neighbour bonds. 
The relation of the magnetic moment on each bond to the magnetic moment of the hexagon centre is constrained by the symmetries of the model, and the appropriate correspondence is shown in Fig.~\ref{fig:honeycomb}(a). 

It is actually very natural for this type of non-local order parameter to arise in the bilayer graphene system.
Given that charge is concentrated at the centre of the honeycomb plaquette \cite{andrei11}, the system must be described by the Wannier orbitals that decay very slowly with the distance from their centre, resulting in an exceptionally large overlap between distant sites \cite{Fu218}. 

We note that the symmetry arguments presented above and detailed in Appendix~\ref{NoLocalOrderParameter} also imply that a symmetry-preserving representation of the 120$^\circ$ coplanar state cannot give rise to a finite spin moment on the honeycomb lattice sites.
We have checked explicitly for the states we construct that the on-site spin expectation values vanish.

In the following, we first consider a honeycomb lattice model with two orbitals per site in Section~\ref{sec:HoneycombQuarterFilling} before turning to a honeycomb model without orbital degrees of freedom in Section~\ref{sec:HoneycombHalfFilling}.
For both models, we focus on a situation where the order parameter represents a true breaking of symmetry.
While we do not explicitly address the possibility below, this generalizes to a scenario in which the order parameter is quantum fluctuating, exactly as discussed at the end of Section~\ref{TriangularLatticeModel} and in Appendix~\ref{app:topo}.

\subsection{Honeycomb model with orbitals at quarter-filling}\label{sec:HoneycombQuarterFilling}

In this section, we obtain an insulating state at quarter-filling using the honeycomb model proposed by
Refs.~\citenum{Fu18} and~\citenum{Vafek18}.
We begin by describing their model before discussing different magnetic bond order parameters and their effects.

Refs.~\citenum{Fu18} and~\citenum{Vafek18} obtain their model by studying the symmetry transformation properties of Bloch functions and comparing against numerical calculations of the microscopic system \cite{Moon12,Nam17,Fu218}.
Since their starting point is the microscopic model, they do not assume the full symmetries of the triangular lattice (Eq.~\eqref{eq:LS}).
In particular, instead of $C_6$, they study the action of $C_3$:
\begin{align}
    C_3:\quad (r_1,r_2)\to(r_1+r_2,-r_1).
\end{align}
This is in agreement with the symmetries shown in the inset of  Fig.~\ref{fig:TBGLatticeGeometry}.

Both Refs.~\citenum{Fu18} and~\citenum{Vafek18} conclude that two orbitals distinguished by their angular momentum in the $z$-direction, $L_z=\pm1$, should lie at each point of the honeycomb lattice.
We label fermions as $c_{\mu,\pm,\br}$, where $\mu=A,B$ specifies the sublattice and `$\pm$' specifies the $L_z$ eigenvalue.
The unit cell position is given by $\br=r_1\veo+r_2\vet$, $r_{1,2}\in\mathds{Z}$, which we take to lie at the centre of the honeycomb plaquettes (shown as black squares in Fig.~\ref{fig:honeycomb}(b)).
The fermion $c_{\mu,\pm,\br}$ is located at $\br+\vec{\tau}_\mu$, $\mu=A,B$ where $\vec{\tau}_\mu$ is defined in Fig.~\ref{fig:honeycomb}(b).
Note that the spin index is suppressed; as above, we assume these indices are acted on by the Pauli matrices $\sigma^\ell$.
Finally, it is convenient to use a vector notation, $c_{\mu,\br}=\left(c_{\mu,+,\br},c_{\mu,-,\br}\right)^T$, $\mu=A,B$ and let the Pauli matrices $\tau^\ell$ act on this pseudospin space.
Under a $C_3$ rotation, the fermions transform as
\begin{align}
    C_3:\quad
    c_{A,\br}&\to 
    e^{2\pi i\tau^z /3}  c_{A,\br'-\veo+\vet},
    &
    c_{B,\br}&\to
    e^{2\pi i \tau^z/3} c_{B,\br'-\veo}
\end{align}
where $\br=(r_1,r_2)$ and $\br'=(r_1+r_2,-r_1).$ 
It follows that the pseudospin symmetry generated by $\tau^\ell$ is not an on-site symmetry of the model, but is intertwined with the rotational symmetry in a nontrivial manner. 

The microscopic provenance of these two orbitals is as follows: $c_{A,+,\br}$ and $c_{B,+,\br}$
originate from fermions near the Dirac point $\vK_\mathrm{lbz}$, while $c_{A,-,\br}$ and $c_{B,-,\br}$ originate from fermions near the Dirac point  $\vK'_\mathrm{lbz}$ \cite{Fu18}.
As we discussed at the beginning of Section~\ref{TriangularLatticeModel}, scattering between fermions from different valleys on different layers involves a very large momentum transfer.
Intervalley hopping terms are therefore  assumed to be small in relation to the other terms of the Hamiltonian.

We now discuss the magnetic insulating state.
Including the sublattice, orbital, and spin degrees of freedom, a total of eight fermions can occupy the unit cell, so that the insulator with electron filling $n_f=2$ should occur at quarter-filling.
We start with a generic hopping Hamiltonian $H^\smHex_t=H_1+H_2+H_3$ where $H_1$, $H_2$, and $H_3$ correspond to nearest, next-nearest, and third-nearest neighbour hopping terms respectively: 
\begin{align}\label{eqn:HoppingHam}
    H_1&=-t_1\sum_\br \Big( c_{A,\br}^\dag c_{B,\br} + c_{A,\br}^\dag c_{B,\br-\veo} + c_{A,\br}^\dag c_{B,\br-\vet}+\Hc\Big)
    \notag\\
    H_2&=-\frac{t_2}{2}\sum_\br\sum_{\mu=A,B}\Big(
    c_{\mu,\br}^\dag c_{\mu,\br-\veo}+c_{\mu,\br}^\dag c_{\mu,\br+\vet}+c_{\mu,\br}^\dag c_{\mu,\br+\veo-\vet} +\Hc\Big)
    \notag\\
    H_3&=-t_3\sum_\br \Big(
    c_{A,\br}^\dag c_{B,\br-\veo-\vet}+c_{A,\br}^\dag c_{B,\br+\veo-\vet}+c_{A,\br}^\dag c_{B,\br-\veo+\vet}+\Hc\Big).
\end{align}
Our energy scale in this section and the next is set by $t_1=1$.
It turns out that $H_2$ will not have a qualitative effect on our results and so we set $t_2=0$ for the remainder of this section. 
The resulting band structure captures the essential features of the microscopic band structure calculations, including the Dirac points at $\vK$ and $\vK'$ and extrema at the $\Gamma$-point. 

In Appendix~\ref{app:QuarterFillingDegen} we show that it is not sufficient to only consider magnetic form factors which are diagonal in the orbital indices.
The symmetry constraints of the coplanar magnetic order imply that a gap cannot be induced at this filling without orbital mixing. 
This is in contrast to the triangular lattice model in which an insulator was obtained without any orbital mixing.
We therefore must consider orbital-mixing order parameters. 
For practical purposes, we limited our  study to bonds no further than third nearest-neighbour apart. 
With this constraint, among the many possible form factors which preserve both $C_3$ and $C_{2y}$, only one leads to a gap at quarter-filling:
\begin{align}\label{eqn:MagneticHam}
    H^\smHex_{\mathrm{mag}}&=P_2
    \sum_\br \bm{\mathcal{P}}(\vK\cdot\br)\cdot\Big(
    c_{A,\br-\veo+\vet}^\dag \tau^x\bm{\sigma} c_{A,\br}
    +c_{A,\br}^\dag \tau_{2\pi/3}\bm{\sigma} c_{A,\br+\vet}
    +c_{A,\br+\vet}^\dag \tau_{4\pi/3}\bm{\sigma} c_{A,\br-\veo+\vet}
\notag\\
    &\quad+c_{B,\br}^\dag \tau^x \bm{\sigma} c_{B,\br-\veo+\vet}+c_{B,\br-\veo+\vet}^\dag \tau_{2\pi/3}\bm{\sigma} c_{B,\br-\veo}+c_{B,\br-\veo}^\dag \tau_{4\pi/3}\bm{\sigma} c_{B,\br} +\Hc\Big)
\end{align}
where $\tau_\theta$ is shorthand for $\tau^x\cos\theta + \tau^y\sin \theta$.
Here, the magnetic moment is positioned on next-nearest neighbour bonds and points in the direction indicated by the diagram in Fig.~\ref{fig:honeycomb}(a).
The bond direction dependent nature of the inter-orbital coupling is a consequence of the non-trivial action of the $C_3$ symmetry on the orbital space. 
While $H^\smHex_\mathrm{mag}$ breaks the pseudospin symmetry generated by $\tau^\ell$, it preserves the physical symmetries of the model.

Since $H^\smHex_\mathrm{mag}$ connects different orbitals, this term involves tunneling \emph{between} the different valleys of the microscopic model on different layers, which we originally assumed was negligible.
However, as the band dispersion close to the magic angle is very flat, there can be a substantial enhancement of the interactions discussed in Section~\ref{TriangularLatticeModel}, leading to the magnetic order we consider.

\begin{figure}
    \centering
    \includegraphics[width=0.99\textwidth]{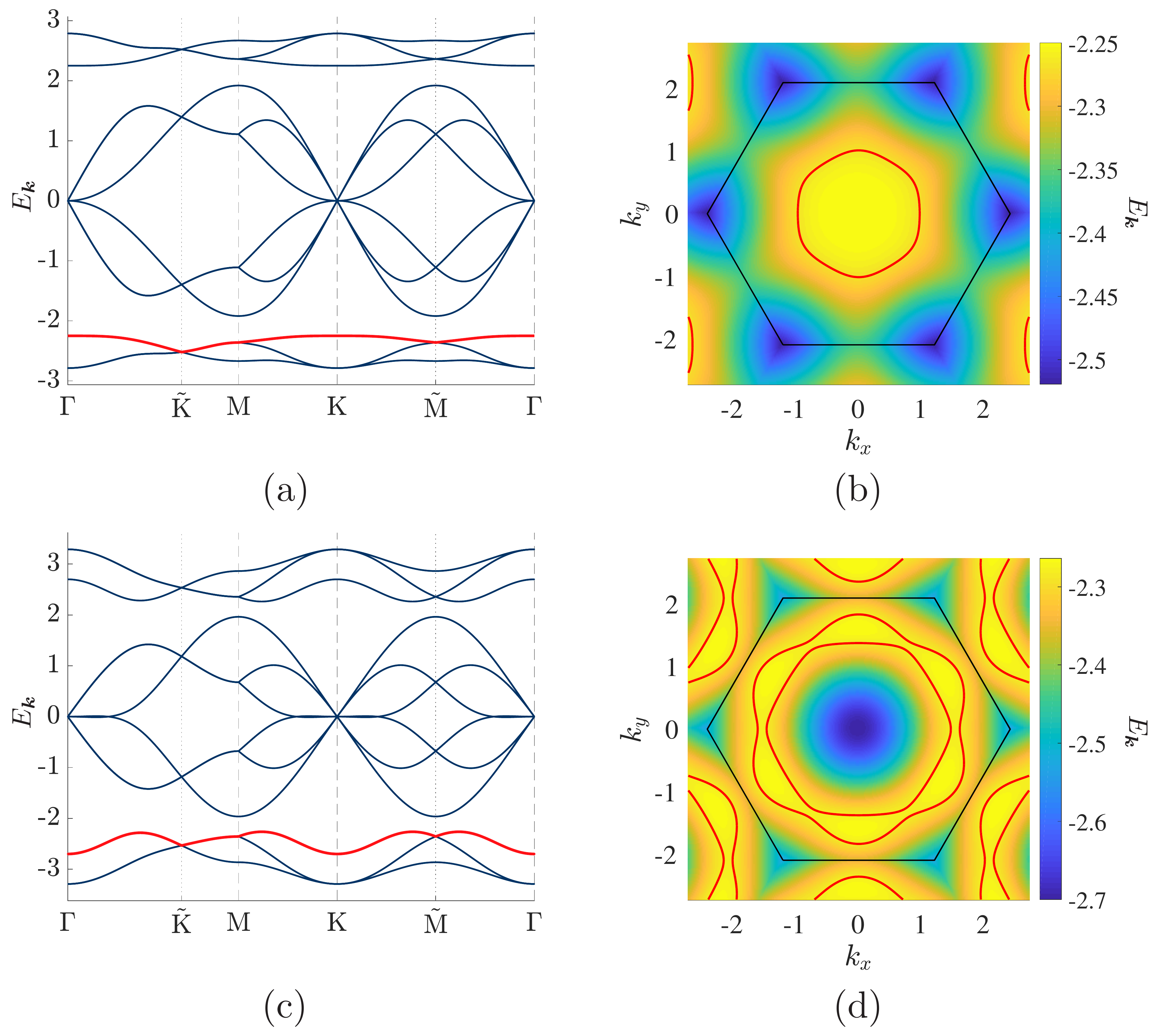}
    \caption{
    (a), (b) $P_2=0.75$, $t_3=-0.15$. 
    (a) The band structure of the Hamiltonian $H^\smHex_t+H_\mathrm{mag}^\smHex$ along the momentum cut in Fig.~\ref{fig:TriangularOPAndSpectrum}(a). As a result of a residual anti-unitary symmetry, all bands plotted are two-fold degenerate.
    (b) Band structure of the band immediately below the gap at one quarter-filling; this corresponds to the band drawn in red in (a).
    The Fermi surface which is obtained upon hole doping is also drawn in red.
    The magnetic BZ is outlined in black. 
    (c), (d) $P_2=0.9$, $t_3=0$. 
    (c) and (d) show the same information as (a) and (b) but without third-nearest-neighbour hopping. 
    }
    \label{fig:BandStrucOrbitalGap}
\end{figure}

The band structure of $H^\smHex=H^\smHex_t+H^\smHex_{\mathrm{mag}}$ is plotted in Figs.~\ref{fig:BandStrucOrbitalGap}(a) and~(c) for $(P_2,t_3)=(0.75,-0.15)$ and for $(P_2,t_3)=(0.9,0.0)$ respectively. 
For both parameter sets, a gap at quarter-filling is clearly visible above the red-coloured band.
As a result of some accidental symmetries, all bands in this model are two-fold degenerate and the spectrum is even about the chemical potential (which is zero in Figs.~\ref{fig:BandStrucOrbitalGap}(a) and~(c)); we discuss this briefly in Appendix~\ref{app:MagSymmetries}.

The Fermi surfaces that are obtained by doping below quarter-filling (in particular, both correspond to a filling $n_f=1.87$) are shown in Figs.~\ref{fig:BandStrucOrbitalGap}(b) and~(d). 
From Fig.~\ref{fig:BandStrucOrbitalGap}(b), we see that a single, doubly-degenerate Fermi surface results when $P_2=0.75$ and $t_3=-0.15$.
This is in agreement with the Shubnikov-de Haas (SdH) oscillations observed below quarter-filling in Refs.~\citenum{Pablo1} and~\citenum{Pablo2}.
Additional terms in the Hamiltonian are allowed which will remove the degeneracy of the Fermi surfaces, but, provided the breaking is not too large, an SdH measurement is unlikely to distinguish the difference in area.

Conversely, when $P_2=0.9$ and $t_3=0$, \emph{two} concentric, doubly-degenerate Fermi surfaces are present, as shown in Fig.~\ref{fig:BandStrucOrbitalGap}(d).
This case is not consistent with the measurements of \refscite{Pablo1,Pablo2}.
Moreover, for larger values of $P_2$ (or smaller filling fractions), these Fermi surfaces break up into six smaller ones.

For both parameter choices, above quarter-filling, there are three electron pockets, similar to what is found on the triangular lattice and shown in \figref{fig:TriangularOPAndSpectrum}(d).

We have also studied the consequences of adding $H^\smHex_\mathrm{mag}$ to the effective tight-binding Hamiltonian detailed in Ref.~\citenum{Fu218} (we included up to fifth nearest-neighbour hoppings).
While this term does induce a gap at quarter-filling when $P_2$ is sufficiently strong, we do not obtain a single Fermi surface on the hole-doped side. 
It is not too surprising that our simple magnetic order parameter is insufficient in the presence of significant hopping between distant sites; when this is the case, magnetic bond order should also be present on these further-neighbour bonds.
It would be interesting to study this further in a self-consistent fashion.
%

\subsection{Honeycomb model at half-filling}\label{sec:HoneycombHalfFilling}
In this section, we study the 120$^\circ$ coplanar antiferromagnet in the context of the model of Ref.~\citenum{Po18}.
We work with a honeycomb model without orbital index and study the state that occurs at half-filling.
We conclude that the 120$^\circ$ magnetic order by itself cannot describe the insulating state.
When magnetic order coexists with valence bond solid order, an insulating state with the desired hole structure is obtained.

The approach of Ref.~\citenum{Po18} differs from that of Refs.~\citenum{Fu18} and~\citenum{Vafek18} in several respects. 
First, the authors of Ref.~\citenum{Po18} argue that, provided the twist angle is small, the symmetries of the triangular lattice (given in Eq.~\eqref{eq:LS}) are approximately correct. 
In particular, they assume a $C_6$ symmetry instead of $C_3$.
This choice is exact for commensurate angles provided the two graphene sheets are rotated about the hexagon \emph{centers} as opposed to Fig.~\ref{fig:TBGLatticeGeometry} where the sheets have been rotated about the honeycomb site.
It is argued in Refs.~\onlinecite{Po18,Senthil2} that the $C_6$ symmetry is approximately preserved regardless of the precise angle or centre of rotation.
Further, since intervalley scattering is assumed to be negligible, they note that global phase transformations may be performed independently on fermions originating from different valleys, and that this implies an emergent U(1)$_v$ valley symmetry.
Topological constraints prevent them from constructing a tight-binding model that preserves both the lattice and U(1)$_v$ symmetries.
As a result, the model they present does not possess the U(1)$_v$ symmetry; it is only obtained through a non-local projection.

We do not work with the fully symmetric tight-binding model of Ref.~\citenum{Po18}, but instead assume that the U(1)$_v$ valley symmetry is spontaneously broken, resulting in what the authors of Ref.~\citenum{Po18} term `intervalley coherent' (IVC) order.
They argue that this is the natural starting point for a study of the insulating state at $n_f=2$.
No topological obstructions remain once the U(1)$_v$ symmetry has been broken, and we assume that a generic tight-binding model on the honeycomb lattice is a sufficient description.
Half of the degrees of freedom are gapped out by the IVC order, and, in contrast to the previous section's model, the appropriate lattice Hamiltonian should only have a \emph{single} orbital per lattice site.
Similar to the triangular lattice model of Section~\ref{TriangularLatticeModel}, the insulating state of interest occurs at \emph{half}-filling.

With the exception of the now-absent orbital index, we use the same notation as in the previous section.
As above, the hopping Hamiltonian is $H^\smHex_t=H_1+H_2+H_3$, where $H_{1,2,3}$ are provided in Eq.~\eqref{eqn:HoppingHam}.

\begin{figure}
    \centering
    \includegraphics[scale=0.35]{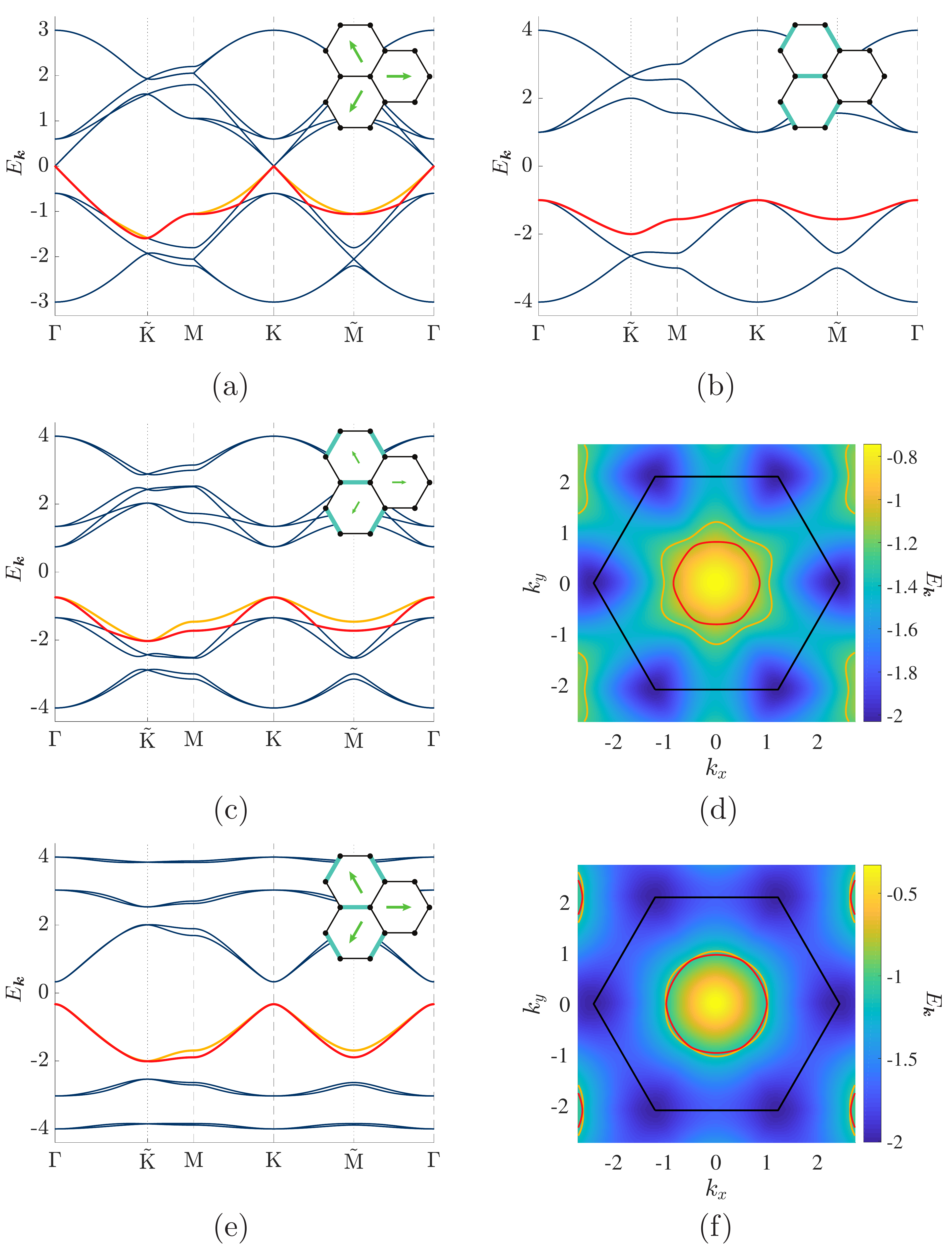}
    \caption{
    (a), (b), (c), and (e) plot the band structure along the cut in Fig.~\ref{fig:TriangularOPAndSpectrum}(a) for various values of $P_1$ and $V$. 
    The orders present are shown in the inset on the top right.
    (d) and (f) show the Fermi surfaces which are obtained by hole-doping the models of (c) and (e), respectively. 
    These are superposed on a color plot of the upper band (coloured orange).
    (a) $V=0.0$, $P_1=0.2$. 
    (b) $V=2.0$, $P_1=0.0$. Here, all bands have a two-fold spin degeneracy.
    (c), (d) $V=2.0$, $P_1=2.0$.  
    The red and orange bands contribute the red and orange Fermi surfaces in (d). 
    (e), (f) $V=2.0$, $P_1=0.9$. The same as shown in (c) and (d), but with a larger magnetic moment.
    }
    \label{fig:VBSBandStruc}
\end{figure}

We now consider the effect at half-filling of adding magnetic bond order to $H_t^\smHex$.
A natural first attempt is to place the order on nearest-neighbour bonds:
\begin{align}
    H_\mathrm{mag}^{\smHex\,\prime}=P_1\sum_\br\bm{\mathcal{P}}(\vK\cdot\br)\cdot\Big(
    c_{A,\br-\veo}^\dagger \bm{\sigma} c_{B,\br-\veo} 
    +
    c_{A,\br}^\dagger \bm{\sigma} c_{B,\br-\vet}
    +
    c_{A,\br+\veo}^\dagger\bm{\sigma}c_{B,\br} +\Hc\Big).
\end{align}
Inspection of Fig.~\ref{fig:honeycomb} shows that this term satisfies the necessary symmetries.
Unfortunately, this order parameter cannot induce a gap at half-filling.
For example, in Fig.~\ref{fig:VBSBandStruc}(a), we plot the band structure of $H^\smHex_t+H^{\smHex\,\prime}_\mathrm{mag}$ with $P_1=0.2$ and $t_2=t_3=0$ (as in the previous section, the energy scale is set by $t_1=1$).
Clearly, no gap is induced, as the Dirac cone at $K$ is still present.
In fact, similar to Section~\ref{sec:HoneycombQuarterFilling}, no choice of order parameter for the 120$^\circ$ antiferromagnetic phase can induce a gap at half-filling; we prove this in Appendix~\ref{app:HalfFillingDegen}.



Conversely, it has been shown that Kekul\'{e} valence band solid (KVBS) order is capable of fully gapping the honeycomb band structure at half-filling \cite{Lee18}.
The KVBS order parameter is illustrated in the inset on the top right of Fig.~\ref{fig:VBSBandStruc}(b).
We see that, like the $120^\circ$ antiferromagnet, KVBS order breaks the translational symmetry and enlarges the unit cell to three hexagons. 
A mean-field representation of this order is obtained by coupling fermions at momentum $\vK$ and $-\vK$:
\begin{align}
    H^\smHex_\mathrm{vbs}&=-\frac{V}{3}\sum_\br\Bigg[
    \left(\cos\left(\vK\cdot2\br\right)+\frac{1}{2}\right)c_{A,\br}^\dagger c_{B,\br} +
    \left(\cos\left(\vK\cdot[2\br-\veo]\right)+\frac{1}{2}\right)c_{A,\br}^\dag c_{B,\br-\veo}
    \notag\\
    &\quad    + 
    \left(\cos\left(\vK\cdot[2\br-\vet]\right)+\frac{1}{2}\right)c_{A,\br}^\dag c_{B,\br-\vet}\Bigg].
\end{align}
(The factor of $1/2$ is added to substract off a constant piece that would otherwise contribute to $H_1$.)
Since $\vK\cdot 2\br = -\vK\cdot\br\text{ mod }2\pi$, it is clear that the band structure should be calculated in the same reduced BZ as illustrated in Fig.~\ref{fig:TriangularOPAndSpectrum}(a). 
In Fig.~\ref{fig:VBSBandStruc}(b) we plot the band structure of $H^\smHex_t+H^\smHex_\mathrm{vbs}$ with $V=2.0$ and $t_2=t_3=0$; the gap at half-filling is clearly visible.

The possibility that KVBS order is present  in bilayer graphene is further supported by quantum Monte Carlo (QMC) simulations.
Ref.~\citenum{Lee18} showed that within a certain parameter regime this order is in fact favoured when the interactions appropriate for the bilayer graphene system are included.
While promising, they find two doubly-degenerate hole pockets below half-filling; this has twice the number of degrees of freedom needed  to describe the SdH oscillations \cite{Pablo1,Pablo2}.
This discrepancy with experiment is also apparent in the mean-field band structure plot of Fig.~\ref{fig:VBSBandStruc}(b).
Further mean-field calculations indicate that a \emph{single} doubly-degenerate Fermi surface is obtained when 
$t_3$ is large and negative ($t_3\lesssim-1.1$), but this is not seen in the QMC  simulation. 

A mean-field band structure with a single pocket can also be obtained by turning on a next-nearest neighbour hopping term, but it must also be large, of the order of $t_1$: $t_2\lesssim-0.8$. 
Since this term breaks the particle-hole symmetry, the model has a sign problem and cannot be studied using QMC.



We show that a state with coexisting KVBS order and magnetic order located on nearest-neighbour bonds can return a single hole-pocket without requiring $|t_2|$ or $|t_3|$ large -- in particular, we let $t_2=t_3=0$.
Figs.~\ref{fig:VBSBandStruc}(c) and~(d) show the spectrum of $H^{\smHex\,\prime}=H^\smHex_t+H^{\smHex\,\prime}_\mathrm{mag}+H^\smHex_\mathrm{vbs}$ when $P_1=0.2$ and $V=2.0$.
The two bands below half-filling are plotted in red and orange in the momentum cut plot in Fig.~\ref{fig:VBSBandStruc}(c), while the Fermi surfaces obtained by hole-doping are plotted in Fig.~\ref{fig:VBSBandStruc}(d) using the same colours as in (c).
They demonstrate that even a relatively modest magnetic moment is sufficient to remove the unwanted pair of Fermi surfaces. 
While the two Fermi surfaces are not identical, distinguishing this scenario from the case of two exactly degenerate Fermi surfaces is difficult to establish by quantum oscillations.
Moreover, as the magnetic moment is increased, the two Fermi surfaces approach one another.
When $P_1=0.9$ and $V=2.0$, the two bands and the corresponding Fermi surfaces are nearly degenerate, as shown in Figs.~\ref{fig:VBSBandStruc}(e) and~(f).

\section{Conclusions}

We have presented the electronic structures of antiferromagnetically ordered states in twisted bilayer graphene. As the electronic charge density takes the form of a triangular lattice \cite{andrei11}, and antiferromagnetism is primarily due to local Coulomb repulsion between the electrons, we only considered states with the symmetry of the 120$^\circ$ coplanar antiferromagnetic order of the triangular lattice \cite{wu2018emergent}. 
Previous studies have determined the electronic structure of such states for electrons
in tight-binding models on the triangular lattice \cite{Tremblay95}. We showed here that the same order can also appear on tight-binding models on the honeycomb lattice,
by allowing for momentum-dependent form factors in the magnetic moments so that the spin density is centered on the bonds of the honeycomb lattice. 
This mechanism is similar to the bond-centered charge density waves considered in the context of the cuprates \cite{SSLR13,Fujita14,ATSS15}. The usual antiferromagnetic state on the honeycomb lattice \cite{Lee18,Ma18} has two sublattices, and the spin density is centered on the sites; in contrast, the state we considered had vanishing spin density on the sites of the honeycomb lattice.

Cao {\em et al.} \cite{Pablo1,Pablo2} have observed quantum oscillations in the hole-doped metal away from the correlated insulator. The electronic orbitals relevant to this observation are in Figs.~\ref{fig:TriangularOPAndSpectrum}(b) and~(c) for the triangular lattice, and in Fig.~\ref{fig:BandStrucOrbitalGap} for the honeycomb lattice. There are parameters with a single doubly-degenerate Fermi surface centered at the $\Gamma$ point, which is consistent with observations.

In Section~\ref{sec:HoneycombHalfFilling}, we also considered cases where the antiferromagnetic order co-exists with intervalley coherent and/or valence bond solid order on the honeycomb lattice and demonstrate that Fermi surfaces consistent with experiment can be obtained.

Our results can also be extended to cases where the antiferromagnetically ordered is not long-ranged, but is quantum fluctuating in a state with $\mathbb{Z}_2$ ({\em i.e.}\/ toric code) topological order. As discussed in Appendix~\ref{app:topo}, such states have fractionalized fermionic `chargons' which inherit the electronic Fermi surfaces of the states described in the main part of the paper, and so will exhibit similar quantum oscillations.

\section*{Acknowledgements}

We thank Shiang Fang, Liang Fu, T.~Kaxiras, T.~Senthil, A.~Vishwanath, Cenke Xu and Liujun Zou for valuable discussions. 
This research was supported by the NSF under Grant DMR-1664842. Research at Perimeter Institute is supported by the Government of Canada through Industry Canada and by the Province of Ontario through the Ministry of Research and Innovation. SS also acknowledges support from Cenovus Energy at Perimeter Institute. 
MS acknowledges support from the German National Academy of Sciences Leopoldina through grant LPDS 2016-12.

\appendix



\section{Quantum fluctuating antiferromagnetism}
\label{app:topo}

This appendix briefly reviews the generalization of antiferromagnetically ordered states to cases where the antiferromagnetism is `fluctuating' and spin rotation invariance is preserved. Provided certain topological defects in the antiferromagnetic order are suppressed, the resulting `quantum disordered' state has $\mathbb{Z}_2$ topological order (similar to that in the `toric code') and preserves gapless fermionic excitations along the Fermi surfaces  described in the body of the paper \cite{SS09,CSS17,Scheurer18}. 

With the aim of placing this discussion in the wider context of the extensive studies of spin liquids, and doped spin liquids, in theories of the cuprates, it is useful to introduce some formalism which places the spin and Nambu pseudospin rotations on an equal footing. To this end, we introduce the matrix fermionic operator
\beq
C_\r = \left(
\begin{array}{cc}
c_{\r \uparrow} & - c_{\r \downarrow}^\dagger \\
c_{\r \downarrow} & c_{\r \uparrow}^\dagger
\end{array}
\right) \label{Xc}
\eeq
We will drop the valley index in this appendix; all the fermions can also carry an implicit valley index. Global spin rotations (denoted here SU(2)$_s$) act via {\it left}-multiplication on $C_\r$ by a SU(2) matrix. Similarly, global Nambu pseudospin rotations (denoted here SU(2)$_c$)
act via {\it right}-multiplication on $C_\r$ by a SU(2) matrix. Note that the pseudospin rotations about the $z$ axis
correspond simply to the U(1) charge conservation symmetry. The full SU(2)$_c$ rotation is not a symmetry of the Hamiltonians we have considered. 

One common way to introduce exotic states with fractionalization is to transform to a rotating reference frame in pseudospin space. This is accomplished by writing \cite{LeeWenRMP,Wen02,XS10}
\beq
C_\r = F_\r R_{\r c} \label{FR}
\eeq
where the $F$ are fermionic spinons defined as in Eq.~(\ref{Xc})
\beq
F_\r = \left(
\begin{array}{cc}
f_{\r \uparrow} & - f_{\r \downarrow}^\dagger \\
f_{\r \downarrow} & f_{\r \uparrow}^\dagger
\end{array}
\right) \label{Ff}
\eeq
while $R_{\r c}$ is a c-number SU(2) matrix. This formulation has `gauged' SU(2)$_c$ to SU(2)$_{cg}$. Under the SU(2)$_{cg}$ generated by $U_\r$, the field transformations are 
\beq
{\text SU(2)}_{cg} : \quad C_\r \rightarrow C_\r, \quad F_\r \rightarrow  F_\r U_\r , \quad R_{\r c} \rightarrow U^\dagger_{\r} R_{\r c} \,,
\eeq
while the global SU(2)$_c$ is
\beq
{\text SU(2)}_{c} : \quad C_\r \rightarrow C_\r \, U, \quad F_\r \rightarrow  F_\r  , \quad R_{\r c} \rightarrow R_{\r c} \, U \,.
\eeq
Note that the SU(2)$_{cg}$ gauge invariance is exact, even though the Hamiltonian is not invariant under SU(2)$_c$. The resulting SU(2)$_{cg}$ gauge theory for $F$ and $R$ describes various insulating spin liquid states in its deconfined states, while confining phases acquire various broken symmetries \cite{LeeWenRMP,Wen02,Wang18,CQSS16,SSSC17,Thomson18}. The doped spin liquids can describe exotic metallic states, provided the $R_{\r c}$ bosons remain uncondensed. Because the $R_{\r c}$ carry U(1) charge, the non-zero charge density requires a non-zero temperature to keep the bosons uncondensed, at least in the most natural mean field theories. The metallic states can also acquire Fermi surfaces of the $F$ spinons, but there is no simple relationship between the volumes enclosed by the spinon Fermi surfaces and doping density, because 
\beq
C_\r^\dagger C_\r \neq F_\r^\dagger F_\r \,. 
\eeq
This feature makes such metallic spin liquids unattractive for graphene. 

The other approach, which has a direct connection between Fermi surface volume and doping density, is obtained by 
transforming to a rotating reference frame in spin space. This is accomplished by writing instead \cite{SS09,CSS17,Scheurer18}
\beq
C_\r =  R_{\r s} \Psi_\r \label{RPsi} \,,
\eeq
where the $\Psi_\r$ are fermionic {\it chargons} defined as in Eq.~(\ref{Xc})
\beq
\Psi_\r = \left(
\begin{array}{cc}
\psi_{\r +} & - \psi_{\r -}^\dagger \\
\psi_{\r - } & \psi_{\r +}^\dagger
\end{array}
\right) \,, \label{Psipsi}
\eeq
while $R_{\r s}$ is a c-number SU(2) matrix. This formulation has now gauged SU(2)$_s$ to SU(2)$_{sg}$. Under the SU(2)$_{sg}$ generated by $U_\r$, the field transformations are 
\beq
{\text SU(2)}_{sg} : \quad C_\r \rightarrow C_\r, \quad \Psi_\r \rightarrow U_\r \Psi_\r  , \quad R_{\r s} \rightarrow  R_{\r s} U^\dagger_{\r} \,,
\eeq
while the global SU(2)$_s$ is
\beq
{\text SU(2)}_{s} : \quad C_\r \rightarrow U C_\r , \quad \Psi_\r \rightarrow  \Psi_\r  , \quad R_{\r s} \rightarrow U R_{\r c}  \,.
\eeq
Now a theory of bosonic spinons, $R_{\r s}$, and fermionic chargons $\Psi_\r$, yields the metallic states with fluctuating antiferromagnetism we wish to describe. Because now 
\beq
C_\r^\dagger C_\r = \Psi_\r^\dagger \Psi_\r\,, 
\eeq
the doping density is directly connected to the volumes enclosed by the $\Psi_\r$ Fermi surfaces.

An important feature of the effective theory of the fermionic chargons, $\Psi_\r$, is the presence of a condensate in a Higgs field, ${\bm H}$. This Higgs field is simply the antiferromagnetic order parameter transformed to the rotating reference under Eq.~(\ref{RPsi}). If we denote the on-site spin moment on the triangular lattice site by ${\bm S}_\r$, then the corresponding on-site Higgs field is 
\beq
{\bm \sigma} \cdot {\bm H}_\r = R_{\r s}^\dagger {\bm \sigma} \cdot {\bm \S}_\r \,R_{\r s}\,.
\eeq
There is a natural generalization of this rotation to the inter-site bi-local moments, ${\bm S}_{\r \r'}$, defined on the links in our paper, and this generalizations yields a corresponding bi-local Higgs field on the same links: 
\beq
{\bm \sigma} \cdot {\bm H}_{\r, \r'} = R_{\r s}^\dagger {\bm \sigma} \cdot {\bm \S}_{\r, \r'} \, R_{\r' s}\,.
\eeq
The transformation of ${\bm H}_\r$ is 
\beq
{\text SU(2)}_{sg} : \quad {\bm \sigma} \cdot {\bm H}_\r \rightarrow U_\r \,{\bm \sigma} \cdot {\bm H}_\r\, U_\r^\dagger\,,
\eeq
with the obvious generalization to bi-local Higgs fields.
The needed states with $\mathbb{Z}_2$ topological order are obtained when
$\left \langle {\bm S}_\r \right\rangle = 0$, but $\left \langle {\bm H}_\r \right\rangle \neq 0$ with precisely the same spatial pattern as the antiferromagnetic order. Because the pattern of the $\left \langle {\bm H}_\r \right\rangle$ is non-collinear, the SU(2)$_{sg}$  is broken down to $\mathbb{Z}_2$. With no continuous gauge invariance remaining unbroken, the fluctuations of the gauge fields are suppressed, and can be safely ignored.
The corresponding effective Hamiltonians for the chargons, $\Psi_\r$, in such metals with $\mathbb{Z}_2$ topological order have the same spatial structure as that for the electrons $C_\r$ described in the body of the paper.

\section{Symmetry constraints on magnetic order}\label{app:SymCons}
The charge density pattern on the moir\'e honeycomb lattice has the symmetries of a triangular lattice. This motivates us to look for magnetic order parameters which, while breaking the global spin rotation symmetry, preserve all lattice symmetries in the charge sector. In other words, any lattice symmetry operation combined with a global spin rotation should be a symmetry of the Hamiltonian\footnote{Note that a local (site-dependent) spin rotation would modify the hopping term that involves electrons at different sites, and is hence not allowed.}. Spin singlet operators that act on the charge sector would be insensitive to a global spin rotation, and therefore preserve the symmetry. In this appendix, we prove that the only such state is the 120$^\circ$ coplanar antiferromagnet on the triangular lattice (excluding the trivial ferromagnet). 

We first show this for decoupled valley indices, taking a single orbital (or spin) at each site of the triangular lattice. In presence of inter-orbital coupling, like the Hund's coupling in Eq.~(\ref{HamiltonianTr}), the lattice symmetry operation needs to be combined with a global rotation that is identical for the spins of the two different valleys. In this case, the only symmetry-allowed state is the 120$^\circ$ coplanar antiferromagnet for each spin, with the spins on the two orbitals either parallel or antiparallel. We argue that the energetics dictated by the Hund's coupling \cite{Xu18} as well as numerical evidence \cite{Ma18} seem to point towards parallel alignment of the spins on different orbitals at the same site. At the end, we discuss additional possibilities that arise on reducing the $C_6$ symmetry of the triangular lattice to the $C_3$ symmetry of the moir\'e superlattice, as shown in Fig.~\ref{fig:TBGLatticeGeometry}.

If the magnetic order parameter $P_0 \bm{\mathcal{P}}(\r)$ (where $P_0$ is the magnitude and $\bm{\mathcal{P}}$ is a unit vector indicating direction) has a spatially varying magnitude, the same will be true of the expectation value of the SU(2) invariant operator $\langle \S^2(\r) \rangle \sim P_0^2(\r)$. Since this breaks translation symmetry for a spin-singlet operator, we need $P_0$ to be spatially uniform. Hence, we choose $P_0 = 1$ (any fixed value of $P_0$ will work for the subsequent arguments). For any symmetry operation $X$, we need a global spin-rotation $U_X$ such that $U_X X$ is a symmetry of the Hamiltonian. Alternately, we need to find a SO(3) rotation $R(\hat{n}_X, \theta_X)$ that rotates the magnetic order at each site $X[\r]$ to the original pattern at site $\r$:
\beq
 X[\bm{\mathcal{P}}(\r)] =  \bm{\mathcal{P}}({X[\r]}) = R(\hat{n}_X, \theta_X)\bm{\mathcal{P}}(\r)
\eeq
Let us first consider the translation operators. Then we require:
\bea
T_1[\bm{\mathcal{P}}(\r)] &=& \bm{\mathcal{P}}(\r + \veo) = R(\hat{n}_{T_1}, \theta_{T_1})\bm{\mathcal{P}}(\r) \nn
T_2[\bm{\mathcal{P}}(\r)] &=& \bm{\mathcal{P}}(\r + \vet) = R(\hat{n}_{T_2}, \theta_{T_2}) \bm{\mathcal{P}}(\r)
\eea
Further, we know that $\bm{\mathcal{P}}(\r + \veo + \vet) = T_1 T_2 [\bm{\mathcal{P}} (\r)] = T_2 T_1 [\bm{\mathcal{P}} (\r)] $, so the SO(3) rotation matrices corresponding to these two operations must commute. 
\beq
[R(\hat{n}_{T_1}, \theta_{T_1}), R(\hat{n}_{T_2}, \theta_{T_2})] = 0
\eeq
Commuting operators preserve eigenspaces, and the axis of rotation is the only real eigenvector of a generic rotation operator in 3 spatial dimensions. Therefore for the two rotations to commute, they must be rotations about the same axis, i.e, $\hat{n}_{T_1} = \hat{n}_{T_2}$\footnote{$\pi$ rotations about perpendicular axes also commute, but they lead to only ferromagnetic states on imposing point group symmetries. The proof is a bit cumbersome, hence not presented.}. Exploiting the spin-rotation symmetry of the underlying Hamiltonian, we can choose $\hat{n}_{T_1} = \hat{n}_{T_2} = \bm{\hat{z}}$. Then, the general magnetic order configuration is given by (re-defining $\theta_{T_i} = \theta_i$ for notational clarity):
\beq
\bm{\mathcal{P}}(\r) = \bm{\mathcal{P}}(r_1,r_2) =  [R(\hat{n}_{T_1}, \theta_1)]^{r_1} [R(\hat{n}_{T_2}, \theta_2)]^{r_2} \bm{\mathcal{P}}(0,0) = R(\bm{\hat{z}}, r_1 \theta_1 + r_2 \theta_2)  \bm{\mathcal{P}}(0,0) 
\eeq
The magnetic order parameter at the origin, $\bm{\mathcal{P}}(0,0)$, has some component along the rotation axis ($\bm{\hat{z}}$) and some component perpendicular to the rotation axis, which we choose to be along $\bm{\hat{x}}$ without loss of generality. Letting $\bm{\mathcal{P}}(0,0) = \mathcal{P}(\sin \alpha, 0, \cos \alpha)$, we find that the most general magnetic state which preserves translation symmetry is given by a conical spiral, which is a commensurate/incommensurate antiferromagnet in-plane with an out-of-plane ferromagnetic component,
\beq
\bm{\mathcal{P}}(\r) = \left( \sin \alpha \cos(\Q \cdot \r), \sin \alpha \sin(\Q \cdot \r), \cos \alpha \right), \text{ where } \Q \cdot \r = \theta_1 r_1 + \theta_2 r_2.
\label{eq:trans}
\eeq

We now impose the point group symmetries, $C_{6}$ and $C_{2y}$. First, consider the rotation $C_{6}$. Using Eq.~(\ref{eq:LS}), we need a rotation $R(\hat{n}_{C_{6}}, \theta_{C_{6}})$ such that $\bm{\mathcal{P}}(C_{6}[\r]) = R(\hat{n}_{C_{6}}, \theta_{C_{6}}) \bm{\mathcal{P}}(\r) $. If $\cos \alpha \neq 0$, then $\hat{n}_{C_{6}}$ must be $\bm{\hat{z}}$, as any other rotation axis would mix the constant $\bm{\hat{z}}$ component with the spatially varying in-plane components. We then find the following constraint.
\beq 
\theta_1 (r_1 + r_2) - \theta_2 r_1 = \theta_1 r_1 + \theta_2 r_2 + \theta_{C_{6}} \mod (2\pi) , ~ \forall ~ r_1, r_2 \in \mathbb{Z}
\label{eq:C6}
\eeq
One can check that the only solutions to Eq.~(\ref{eq:C6}) are $\theta_1 = \theta_2 = 0$. This solution is the ferromagnet, which trivially preserves all point group symmetries as well, and we ignore it henceforth.

This leads us to consider coplanar states with $\cos\alpha = 0$. $C_{6}$ fixes the origin, so we must have $R(\hat{n}_{C_{6}}, \theta_{C_{6}}) \bm{\mathcal{P}}(0,0) = \bm{\mathcal{P}}(0,0) = \mathcal{P}(1,0,0)$. This constrains the rotation axis to $\bm{\hat{x}}$. By examining any point with $\bm{\mathcal{P}} \cdot \bm{\hat{y}} \neq 0$, we see that $\theta_{C_{6}} = \pi$ is the only angle which preserves the coplanar nature of the ordered state. Imposing $\bm{\mathcal{P}}(C_{6}[\r]) = R(\bm{\hat{x}}, \pi) \bm{\mathcal{P}}(\r)$, we find
\beq
\theta_1 (r_1 + r_2) - \theta_2 r_1 = - (\theta_1 r_1 + \theta_2 r_2), ~ \forall ~ r_1, r_2 \in \mathbb{Z}
\label{eq:C6ip}
\eeq
There only solution to Eq.~(\ref{eq:C6ip}), given by $\theta_1 = -\theta_2 = \pm 2\pi/3$. Thus, imposing $C_{6}$ already restricts us to the 120$^\circ$  coplanar antiferromagnet with wave-vector $\Q = \pm \vK$. 

Finally, consider the reflection $C_{2y}$, for which we need a rotation $R(\hat{n}_{C_{2y}}, \theta_{C_{2y}}) \bm{\mathcal{P}}(\r)$ such that $\bm{\mathcal{P}}(C_{2y}[\r]) = R(\hat{n}_{C_{2y}}, \theta_{C_{2y}}) \bm{\mathcal{P}}(\r)$. We consider a rotation about $\hat{n}_{C_{2y}} = \bm{\hat{z}}$, which gives
\beq
 -(\theta_1 r_2 + \theta_2 r_1) = \theta_1 r_1 + \theta_2 r_2 + \theta_{C_{2y}} \mod 2\pi ~ \forall ~ r_1, r_2 \in \mathbb{Z}
\label{eq:Rx}
\eeq
There is a one-parameter family of solutions to Eq.~(\ref{eq:Rx}), given by $\theta_1 = -\theta_2$ and $\theta_{C_{2y}} = 0$, for arbitrary $\cos\alpha$. However, the solution which preserved $C_{6}$ was a specific member of this family of solutions, as it had $\theta_1 = -\theta_2 = \pm 2\pi/3$ and $\cos\alpha = 0$.  Thus, we conclude that the only state which preserves all symmetries of the triangular lattice for spin-rotation invariant observables is the 120$^\circ$ coplanar antiferromagnet (excluding the ferromagnet). 

Next, we study the case with two orbitals at each site. Without inter-orbital coupling, each orbital can have its own magnetic moment arranged in the 120$^\circ$ coplanar antiferromagnetic pattern on the triangular lattice, with an arbitrary angle between the two moments at the same site. In presence of inter-orbital coupling that conserves the total spin at each site, the only rotation generators allowed are those which rotate the total spin. For $C_6$ rotations, this implies that $R(\bm{\hat{x}},\pi)$ for each moment must be replaced by an appropriate $\pi$-rotation about the same axis for both spins. However, the origin is a fixed point of $C_6$, and therefore such a rotation must not change either of the spins at the origin. Taken together, these imply that the spins in the two orbitals at the origin (and therefore, on every site) must be either parallel or antiparallel. The Hund's coupling in Eq.~(\ref{HamiltonianTr}) favors a parallel alignment, and numerical studies of twisted bilayer graphene in Ref.~\onlinecite{Ma18} observe sizable antiferromagnetic correlations at large $U$, which would not be the case if the spins in the two orbitals at the same site were anti-aligned. Therefore, we conclude that the magnetic moments of the two orbitals must be aligned at each site. 

Finally, it is interesting to note that if we reduce the $C_{6}$ symmetry to $C_3$, then the same conclusion holds if $C_{2x}$ is preserved, where  $C_{2x}$ is $\pi$ rotation about $\bm{\hat{x}}$, which in the effective two-dimensional model corresponds to a reflection about $\bm{\hat{x}}$.
\beq
C_{2x}:\quad(r_1,r_2)\to (r_2,r_1)
\eeq
Such a state preserves both $C_{6}$ and $C_{2y}$, as discussed earlier. However, a non-coplanar 120$^\circ$ \textit{conical} antiferromagnet (of the form of Eq.~(\ref{eq:trans}) with $\theta_1 = - \theta_2 = \pm 2\pi/3$ and $\cos\alpha \neq 0$) is allowed if $C_3$ and $C_{2y}$ are preserved instead, as is the case with the moir\'e superlattice of twisted bilayer graphene (see Fig.~\ref{fig:TBGLatticeGeometry}). Such a state breaks $C_{2x}$ for spin singlet observables, and allows for non-zero on-site (ferromagnetic) moments on the honeycomb lattice sites. However, it does not seem to be energetically favorable in the insulator on the triangular lattice model \cite{JG1989,BLP1992,Deutscher1993}, and an on-site ferromagnetic moment is yet to be observed in numerical studies of the honeycomb lattice model in the context of bilayer graphene \cite{Lee18,Ma18}.

\section{Absence of local order parameter on the honeycomb lattice}\label{NoLocalOrderParameter}
As it is a central aspect of this work, we here provide the complete proof for why an on-site magnetic order parameter cannot describe the 120$^\circ$ coplanar state for any honeycomb lattice model.

To show this statement formally and see why the inclusion of several orbitals per honeycomb site does not affect the result, consider the following general magnetic on-site order parameter
\begin{equation}
    H_{\text{mag}} = \sum_{\br}\sum_{\mu=A,B} \vec{\mathcal{P}}_{\mu,\br} \cdot
    \left(
    \sum_{\ell=0,x,y,z}f_{\mu,\ell}(\br)\, c^\dagger_{\mu,\br} \tau^\ell \vec{\sigma} c^\pdagger_{\mu,\br}
    \right), \label{GeneralOnSiteOrderParameter}
\end{equation}
where $\br$ is summed over the sites of the triangular lattice and $f_{\mu,\ell}(\br)$ is a form factor that allows for arbitrary mixing of the different orbitals, on each honeycomb lattice site.

Consider a fixed site $\br=\br_0$ on the $\mu$ sublattice. As can be seen in \figref{fig:honeycomb}(a), three-fold spatial rotation with axis perpendicular to the 2D plane and through the site $\br_\mu=\br+\vec{\tau}_\mu$ accompanied by a three-fold rotation in spin space  (with same or opposite orientation depending on the sublattice $\mu=A$ or $B$) is a symmetry of the 120$^\circ$ coplanar order parameter and, for commensurate twist angles, also of the twisted bilayer structure in \figref{fig:TBGLatticeGeometry}. 
Invariance of \equref{GeneralOnSiteOrderParameter} requires
\begin{equation}
    \vec{\mathcal{P}}_{\mu,\br_0} \cdot  \left(U^\dagger_{3} \vec{\sigma}U^\pdagger_{3}\right)\sum_\ell f_{\mu,\ell}(\br_0)\,\widetilde{U}^\dagger_{3} \tau^\ell \widetilde{U}^\pdagger_{3}
    = 
    \left(\vec{\mathcal{P}}_{\mu,\br_0} \cdot \vec{\sigma}\right)
    \sum_\ell f_{\mu,\ell}(\br_0)\tau^\ell, \label{ConditionForInvariance}
\end{equation}
where $\widetilde{U}_{3}$ and $U_{3}$ denote the representation of the symmetry operation in orbital and spin space, respectively. 
$U_{3}=e^{\pm i \frac{2\pi}{3}\frac{\sigma^z}{2}}$ with the sign depending on the sublattice as discussed above. 
The explicit form of $\widetilde{U}_{3}$ is irrelevant for the current argument.
A non-zero order parameter in \equref{GeneralOnSiteOrderParameter} requires that at least one of the components of the form factor $f_{\mu,\ell}(\br_0)$ be non-zero.
Suppose we have $f_{\mu,\ell_0}(\br_0)\neq0$.
\equref{ConditionForInvariance}, in turn, requires that   \begin{align}
    \frac{1}{2}\sum_\ell f_{\mu,\ell}(\br_0)
    \mathrm{tr}\left( \tau^{\ell_0} \widetilde{U}^\dagger_{3}\tau^\ell \widetilde{U}^\pdagger_{3}
    \right)
\end{align} 
be non-zero as well, where the trace is taken over the orbital indices.
Multiplying each side of \equref{ConditionForInvariance} by $\tau^{\ell_0}$ and tracing over the orbital indices implies that
\begin{equation}
   R_{3} \vec{\mathcal{P}}_{\mu,\br_0} = \pm C \, \vec{\mathcal{P}}_{\mu,\br_0}, \qquad 
   C = f_{\mu,\ell_0}(\br_0) /\left[\frac{1}{2}
   \sum_\ell f_{\mu,\ell}(\br_0)\mathrm{tr}\left(\tau^{\ell_0}\widetilde{U}^\dagger_{3} \tau^\ell\widetilde{U}^\pdagger_{3}\right)\right] \neq 0, \label{SimplifiedCriterion}
\end{equation}
must hold, where $R_{3}$ rotates vectors by angle $2\pi/3$ about the $z$ axis. 
This is only consistent with $\vec{\mathcal{P}}_{\mu,\br_0}=P_{\mu,\br_0} \hat{\vec{z}}$ (and $C = \pm 1$).

Although consistent with the three-fold rotation symmetry, $\vec{\mathcal{P}}_{\mu,\br_0}=P_{\mu,\br_0} \hat{\vec{z}}$ in \equref{GeneralOnSiteOrderParameter}, is clearly not a representation of the 120$^\circ$ coplanar state in \figref{fig:honeycomb}(a) on the honeycomb lattice: the latter is odd under $\pi$-rotation in spin space along the $\sigma^z$ axis, while the order parameter with $\vec{\mathcal{P}}_{\mu,\br_0}=P_{\mu,\br_0} \hat{\vec{z}}$ in \equref{GeneralOnSiteOrderParameter} will be even under this symmetry operation. 

From the symmetry arguments presented above, it also follows that any symmetry-preserving representation of the 120$^\circ$ coplanar state cannot give rise to a finite spin moment on the honeycomb lattice sites: the three-fold rotation symmetry discussed above forces the in-plane component of the spin expectation value to vanish, while the combination of $\pi$-rotation along $\sigma^z$ and time-reversal, which is a symmetry of 120$^\circ$ coplanar state, leads to a vanishing $z$-component of the spin.

\section{Symmetry constraints on the honeycomb lattice band structure}
\subsection{Symmetry action on fermions}\label{app:FermSyms}

In this appendix, we show how the symmetries (Eq.~\eqref{eq:LS}) act on the fermions in Ref.~\citenum{Fu18} and~\citenum{Vafek18}'s models. We label the fermion operators at each site by their sublattice, $A,B$, as vector with respect to their spin and momentum indices:
\begin{align}
    c_{\mu,\br}&=\left( c_{\mu,+,\uparrow},c_{\mu,+,\downarrow},c_{\mu,-,\uparrow},c_{\mu,-,\downarrow}\right)^T,
    &
    \mu&=A,B.
\end{align}
Pauli matrices $\tau^\ell$ act on the orbital index, $\pm$, while Pauli matrices $\sigma^\ell$ act on the spin index. 
The symmetries act on these operators as
\begin{align}\label{eqn:SymActionOrbitals}
    T_1&:\quad c_{\mu,\br}\to c_{\mu,\br+\veo},
    &
    T_2&:\quad c_{\mu,\br}\to c_{\mu,\br+\vet},
    \notag\\
    C_{2y} &:\quad 
    \begin{pmatrix} c_{A,\br}\\ c_{B,\br} \end{pmatrix}
    \to 
    -\tau^x 
    \begin{pmatrix}c_{B,C_{2y}[\br]}\\c_{A,C_{2y}[\br]} \end{pmatrix},
    &
    C_3&:\quad
    \begin{pmatrix} c_{A,\br}\\ c_{B,\br} \end{pmatrix}
    \to
    -e^{2\pi i\tau^z/3}
    \begin{pmatrix} c_{A,C_3[\br]-\veo+\vet} \\ c_{B,C_3[\br]-\veo} \end{pmatrix}.
\end{align}
It will make sense to treat the sublattice index as a final pseudospin index: $c_\br=(c_{A,\br},c_{B,\br})^T$. We let this index be acted on by Pauli matrices $\eta^\ell$.  


\subsection{Gauge transformation and momentum space represenation}

In the main body of the text, we diagonalize the Hamiltonian in the magnetic BZ shown in Fig.~\ref{fig:BandStrucOrbitalGap}(a);
however, it will be convenient in these appendices to work with gauge transformed fermions
\begin{align}\label{eqn:GaugeTransFermions}
    c_{\mu,\br} = e^{-i\sigma^z \vK\cdot\br/2}\psi_{\mu,\br }.
\end{align}
In terms of these $\psi$-fermions, the magnetic Hamiltonianm $H_\mathrm{mag}$ in Eq.~\eqref{eqn:GeneralOrderParam}, takes the form
\begin{align}
    H_\mathrm{mag}&=\sum_\br \sum_{\substack{\mu,\nu\\\bm{a},\bm{a}'}} f_{\bm{a},\bm{a}'}^{\mu\nu} \psi^\dagger_{\mu,\br+\bm{a}} e^{i\sigma^z \vK\cdot\bm{a}/2}\sigma^x e^{-i\sigma^z\vK\cdot\bm{a}'/2}\psi_{\nu,\br+\bm{a}'},
\end{align}
where $\bm{a}$ and $\bm{a}'$ are Bravais lattice vectors and $f_{\bm{a},\bm{a}'}^{\mu\nu}$ is the form factor.
Note that we have dropped the ``\hexagon" superscript since this appendix deals only with the honeycomb lattice.
We next express a generic hopping Hamiltonian in terms of the $\psi$ fermions:
\begin{align}
    H_t&=\sum_\br \sum_{\substack{\mu,\nu\\\bm{a},\bm{a}'}}t_{\bm{a},\bm{a}'}^{\mu\nu} c_{\mu,\br+\bm{a}}^\dagger c_{\nu,\br+\bm{a}'}
    =
    \sum_\br\sum_{\substack{\mu,\nu\\ \bm{a},\bm{a}'}}t_{\bm{a},\bm{a}'}^{\mu\nu}
    \psi^\dagger_{\mu,\br+\bm{a}}e^{i\sigma^z \vK\cdot(\bm{a}-\bm{a}')/2} \psi_{\nu,\br+\bm{a}'}.
\end{align}
Both $H_M$ and $H_t$ are translationally invariant and can be defined on the entire BZ -- we have eliminated the need to define a magnetic BZ. 
Instead, we can write
\begin{align}
    H&=H_t+H_\mathrm{mag}=
    \int\frac{d^2k}{(2\pi)^2}\,\sum_{\ell=0,x,y,z} h_\ell^{\mu\nu}(\k)\psi_{\mu,\k}^\dagger \sigma^\ell \psi_{\nu,\k},
    \notag\\
    h_\ell^{\mu\nu}(\k) &= \frac{1}{2}\sum_{\bm{a},\bm{a}'}
    s^{\mu\nu}_{\bm{a},\bm{a}'}(\ell)e^{i\k\cdot(\bm{a}-\bm{a}')}\mathrm{tr}\left(\sigma^\ell e^{i\sigma^z \vK\cdot \bm{a}/2}\sigma^x e^{-i\sigma^z \vK\cdot \bm{a}'/2}\right),
\end{align}
where $s^{\mu\nu}_{\bm{a},\bm{a}'}(\ell)$ equals $f^{\mu\nu}_{\bm{a},\bm{a}'}$ for $\ell=x,y$ and $t^{\mu\nu}_{\bm{a},\bm{a}'}$ for $\ell=0,z$ and the trace is performed over the spin indices.
The momenta are expressed in terms of the reciprocal vectors, $\k=k_1 \bm{g}_1+k_2\bm{g}_2$, where $\bm{g}_{1,2}$ are defined such that $\bm{g}_i\cdot \bm{e}_j=\delta_{ij}$.
$k_1$ and $k_2$ are both integrated from $-\pi$ to $\pi$.

Finally, the sublattice degrees of freedom can be written in terms of the Pauli matrices $\eta^\ell$, giving
\begin{align}\label{eqn:GaugeTransHamiltonian}
    H&=\int \frac{d^2k}{(2\pi)^2}\psi_\k^\dagger h(\k) \psi_\k,
    &
    h(\k)=\sum_{n,\ell}h^{n\ell}(\k) \eta^n\sigma^\ell .
\end{align}

\subsection{Degeneracy constraints at quarter-filling}\label{app:QuarterFillingDegen}

In this section, we demonstrate that given the symmetry action in Eq.~\eqref{eqn:SymActionOrbitals}, a gap cannot be induced a quarter-filling when both the magnetic order and hopping Hamiltonian are diagonal in orbital space.

We consider the action of $C_{2y}$ on the Hamiltonian, using the form in Eq.~\eqref{eqn:GaugeTransFermions}.
This acts on the momentum-space $\psi$ fermions as
\begin{align}\label{eqn:RxSymPsiFerm}
    C_{2y}&:\quad\psi_{(k_1,k_2)}\to \eta^x\psi_{(-k_2+\pi,k_1+\pi)}.
\end{align}
Under this transformation, the point $\bm{M}=(0,\pi)$ is mapped to itself.
It follows that in order for this symmetry to be preserved preserved $ h(\bm{M})=\eta^x h(\bm{M})\eta^x$, which implies that
\begin{align}
    h(\bm{M})=\sum_{\ell=0,x,y,z}
    \left( c_{0,\ell}\sigma^\ell + c_{x,\ell}\eta^x\sigma^\ell\right),
\end{align}
where $c_{0,\ell},c_{x,\ell}$ are real numbers.

We next consider the constraints imposed by the $C_3$ symmetry. 
This acts on the $\psi$ fermions as
\begin{align}\label{eqn:C3SymPsiFerm}
    C_3&:\quad
    \psi_{(k_1,k_2)}\to
    e^{-ik_1/2}e^{ik_2}U_3\psi_{(-k_1+k_2+\pi,-k_1+\pi)},
    &
    U_3(\vk)&=e^{ik_1\eta^z/2}e^{i\pi\eta^z\sigma^z/3}e^{i2\pi  \tau^z/3}.
\end{align}
Again, $\bm{M}=(0,\pi)$ is mapped onto itself. 
Using the constraint imposed by $C_{2y}$ symmetry, we conclude that
\begin{align}
    \sum_\ell \left( c_{0,\ell}\sigma^\ell + c_{x,\ell}\eta^x\sigma^\ell\right)
    =
    \sum_\ell \left(
    c_{0,\ell} U_3^\dagger(\bm{M}) \sigma^\ell U_3(\bm{M})+ c_{x,\ell} U_3^\dagger(\bm{M}) \eta^x \sigma^\ell U_3(\bm{M}) \right).
\end{align}
In order for the equality to hold, the Hamiltonian must take the form
\begin{align}
    h(\bm{M})=c_{0,0}\mathds{1}+c_{0,z}\sigma^z+c_{x,x} \eta^x\sigma^x+ c_{x,y}\eta^x\sigma^y.
\end{align}
This has eigenstates $c_{0,0}\pm \sqrt{c_{x,x}^2+c_{x,y}^2+c_{0,z}^2}$ -- proving that a degeneracy cannot be induced at quarter-filling without breaking these symmetries. 

\subsection{Degeneracy constraints at half-filling}\label{app:HalfFillingDegen}

In this section we demonstrate that given the symmetry action in Eq.~\eqref{eqn:SymActionOrbitals} no gap can be induced at 1/2-filling in a model without orbitals.
As above, we work with the gauge-transformed fermions in Eq.~\eqref{eqn:GaugeTransFermions} and the Hamiltnonian in Eq.~\eqref{eqn:GaugeTransHamiltonian}. 
The arguments are all identical to what we saw in the previous section, except at the momentum point $\vK/2=\frac{2\pi}{3}\bm{g}_1+\frac{\pi}{3}\bm{g}_2=(2\pi/3,\pi/3)$.

According to Eq.~\eqref{eqn:RxSymPsiFerm}, $C_{2y}$ maps $\vK/2$ back to itself. 
As above, we conclude that
\begin{align}
    h(\vK/2)
    =\sum_\ell
    \left( c'_{0,\ell}\sigma^\ell + c'_{x,\ell}\eta^x\sigma^\ell\right),
\end{align}
for real numbers $c'_{0,\ell}, c_{x,\ell}'$.
$\vK/2$ is also mapped to itself under the $C_3$ transformation in Eq.~\eqref{eqn:C3SymPsiFerm}, implying that
\begin{align}
    \sum_\ell \left( c'_{0,\ell}\sigma^\ell + c'_{x,\ell}\eta^x\sigma^\ell\right)
    &=\sum_\ell \left( c'_{0,\ell} U_3^\dag(\vK/2) \sigma^\ell U_3(\vK/2)+ c'_{x,\ell}U_3^\dag(\vK/2)\eta^x\sigma^\ell U_3(\vK/2)\right).
\end{align}
It follows that at $\vK/2$ the Hamiltonian must take the form
\begin{align}
    h(\vK/2)=
    c'_{0,0}\mathds{1} + c'_{0,z}\sigma^z + c'_x\eta^x(\mathds{1}-\sigma^z),
\end{align}
where $c'_x=c'_{x,0}=-c'_{x,z}$
Solving, we find doubly degenerate eigenvalues $c'_{0,0}+c'_{0,z}$. 

\section{\texorpdfstring{Symmetries of $H^\smHex_t+H^\smHex_\mathrm{mag}$}{Symmetries of H\textasciicircum hex}}
\label{app:MagSymmetries}

The Hamiltonian $H^\smHex_t+H^\smHex_\mathrm{mag}$ has a number of symmetries in addition to the physical symmetries discussed in the main text. 
It actually more convenient to use a slightly different version of the gauge-transformed fermions: $\psi_\vk=e^{-i\pi\sigma^z/6}\tilde{\psi}_\vk.$
The momentum-space representation in terms of the $\tilde{\psi}_\vk$'s is of the form Eq.~\eqref{eqn:GaugeTransHamiltonian}.

We separate the kernel $h(\vk)$ into two pieces: $h(\vk)=h_t(\vk)+h_\mathrm{mag}(\vk)$ (as above, we suppress the ``\hexagon" superscript).
The magnetic contribution to the Hamiltonian takes the form
\begin{align}
    h_\mathrm{mag}(\vk)&=
    P_0\left(
    f_{0,xx}(\k)\tau^x\sigma^x
    +
    f_{0,yx}(\k)\tau^y\sigma^x
    +
    f_{z,yx}(\k)\eta^z\tau^y\sigma^x
    +
    f_{z,yy}(\k) \eta^z\tau^y\sigma^y 
    \right),
\end{align}
where
\begin{align}
    f_{0,xx}(\vk)&=\cos(k_1-k_2)-\frac{1}{4}(\cos k_1-\cos k_2),
    &
    f_{0,yx}(\vk)&=-\frac{\sqrt{3}}{4}\left(\cos k_1+\cos k_2\right),
    \notag\\
    f_{z,yx}(\vk)&=\frac{\sqrt{3}}{4}\left(\cos k_1 +\cos k_2\right),
    &
    f_{z,yy}(\vk)&=\frac{3}{4}\left(\cos k_1 -\cos k_2\right).
\end{align}
The hopping portion is given by
\begin{align}
    h_t(\vk)&=
    g_{x,0}(\vk)\eta^x
    +
    g_{x,z}(\vk)\eta^x\sigma^z
    +
    g_{y,0}(\vk)\eta^y
    +
    g_{y,z}(\vk)\eta^y\sigma^z
\end{align}
where
\begin{align}
    g_{j,0}(\vk)&=\frac{1}{2}\left( h_j(\vk+\vK/2)+h_j(\vk-\vK/2)\right),
    \notag\\
    g_{j,z}(\vk)&=\frac{1}{2}\left( 
    h_j(\vk+\vK/2)-h_j(\vk-\vK/2)\right),
\end{align}
and
\begin{align}
    h_x(\vk)&=
    -t_1\left( 1+\cos k_1 +\cos k_2\right)
    -t_3\Big( 2\cos(k_1-k_2)+\cos(k_1+k_2)\Big),
    \notag\\
    h_y(\vk)&=
    t_1\left( \sin k_1 + \sin k_2 \right)
    +t_3\sin(k_1+k_2).
\end{align}

By inspection, we note that $[h(\vk),\tau^z\sigma^z]=0$, implying that simultaneous rotations about both the orbital and spin $z$-directions are conserved. 
(This is also clear from the real-space representation in Eqs.~\eqref{eqn:MagOrder} and~\eqref{eqn:HoppingHam}.)
While time-reversal is broken, there is an anti-unitary symmetry under which the sublattice and orbitals indices are interchanged: {$\eta^x\tau^xh^*(\vk)\tau^x\eta^x = h(\vk).$}
Since $\tau^x\eta^x$ and $\tau^z\sigma^z$ anticommute, all bands must be two-fold degenerate, one with $\tau^z\sigma^z=+1$ and the other with $\tau^z\sigma^z=-1$.
In addition, there are two chiral symmetries:
$\{\eta^z\tau^z,h(\vk)\}=\{\eta^z\sigma^z,h(\vk)\}=0$.
These constrain the spectrum to be even about zero (or the chemical potential, when it is present).


\bibliography{graphene}

\end{document}